%% file: 5Gpolar.tex
\documentclass[journal,10pt]{IEEEtran}

\usepackage{cite}
\usepackage{epsfig}
\usepackage{epstopdf}
\usepackage{graphicx}
\usepackage{xcolor}
\usepackage{tikz}
\usetikzlibrary{arrows,positioning,shapes.geometric,circuits.logic.US}
\usepackage{pgfplots}
\usepackage{multirow}
\usepackage{upgreek}
\usepackage{amssymb}
\usepackage{amsmath}
\usepackage{cases}
\usepackage{url}
\usepackage{pifont}
\usepackage{bm}
\usepackage{amsthm}

\usepackage{tabularx}
\usepackage{booktabs}
\usepackage{filecontents}
\usepackage{subcaption}
\newcolumntype{Y}{>{\centering\arraybackslash}X}

\usepackage{algorithm}
\usepackage{algpseudocode}

\usepackage{array}

\newcommand{\fixme}[2]{\ifx&#2&{\leavevmode\color{red}#1}\else{\leavevmode\color{red}FIXME\{}#1{\leavevmode\color{red}\}}\footnote{{\leavevmode\color{red}#2}}\PackageWarning{Fixme}{#1: #2}\fi}

\newcommand{\newstuff}[2]{\ifx&#2&{\leavevmode\color{blue}#1}\else{\leavevmode\color{blue}FIXME\{}#1{\leavevmode\color{blue}\}}\footnote{{\leavevmode\color{blue}#2}}\PackageWarning{Newstuff}{#1: #2}\fi}

\DeclareMathOperator{\modulo}{mod}

\title{Design of Polar Codes in 5G New Radio}

\author{Valerio~Bioglio,~\IEEEmembership{Member,~IEEE,}
		Carlo~Condo,~\IEEEmembership{Member,~IEEE,}
        Ingmar~Land,~\IEEEmembership{Senior~Member,~IEEE}
}

\begin{document}

\maketitle
\begin{abstract}
Polar codes have attracted the attention of academia and industry alike in the past decade, such that the 5$^\text{th}$ generation wireless systems (5G) standardization process of the 3$^\text{rd}$ generation partnership project (3GPP) chose polar codes as a channel coding scheme. 
In this tutorial, we provide a description of the encoding process of polar codes adopted by the 5G standard. 
We illustrate the struggles of designing a family of polar codes able to satisfy the demands of 5G systems, with particular attention to rate flexibility and low decoding latency. 
The result of these efforts is an elaborate framework that applies novel coding techniques to provide a solid channel code for NR requirements. 
\end{abstract}



\IEEEpeerreviewmaketitle

\section{Introduction} \label{sec:intro}

Polar codes are a class of capacity-achieving codes introduced by Ar{\i}kan  \cite{arikan}. 
In the past decade, the interest and research effort on polar codes has been constantly rising in academia and industry alike. 
Within the ongoing 5$^\text{th}$ generation wireless systems (5G) standardization process of the 3$^\text{rd}$ generation partnership project (3GPP), polar codes have been adopted as channel coding for uplink and downlink control information for the enhanced mobile broadband (eMBB) communication service. 
5G foresees two other frameworks, namely ultra-reliable low-latency communications (URLLC) and massive machine-type communications (mMTC), for which polar codes are among the possible coding schemes.

The construction of a polar code involves the identification of channel reliability values associated to each bit to be encoded. 
This identification can be effectively performed given a code length and a specific signal-to-noise ratio. 
However, within the 5G framework, various code lengths, rates and channel conditions are foreseen, and having a different reliability vector for each parameter combination is unfeasible. 
Thus, substantial effort has been put in the design of polar codes that are easy to implement, having low description complexity, while maintaining good error-correction performance over multiple code and channel parameters.

The majority of available literature does not take into account the specific codes designed for 5G and their encoding process; given their upcoming widespread utilization, the research community would benefit from considering them within error-correction performance evaluations and encoder/decoder designs. 
Both the encoding and the decoding process can in fact incur substantial speed and complexity overhead, while the performance of decoders is tightly bound to the characteristics of the polar code. 
Works focusing on hardware and software implementations can effectively broaden their audience by including compliance to the 5G standard. 

An industry standard is a document providing specifications for delivering a service agreed upon by a group of competing companies. 
This agreement allows different manufacturers to create products that are compatible with each other, so that standard details are often the result of a \textit{quid pro quo} among companies. 
The outcome of the endless discussions and struggles among different agendas is a patchwork of techniques, whose mixture provides acceptable performance; for this reason, a standard usually represents the state-of-the-art of a field more than its pinnacle. 

In this paper, we provide a tutorial for the polar code encoding process foreseen by 5G in \cite{3GPP_R15}, from the code concatenation, through interleaving functions, to the polar-code specific subchannel allocation and rate-matching schemes. 
The purpose of this work is to provide the reader with a straightforward, self-contained guide to the understanding and implementation of 5G-compliant encoding of polar codes. 
While this work does not claim to substitute the reading of the standard, we aim at assisting the reader in its comprehension, restructuring its presentation and reformulating some of the contents to improve readability. 

The remainder of the paper is organized as follows. 
Section \ref{sec:prel} introduces the basics on polar codes, while Section \ref{sec:adv} details more advanced design features, along with concepts used in the 5G encoding process, such as interleaving and rate-matching. 
Section \ref{sec:5G} details a step-by-step guide to 5G polar code encoding. 
Decoding considerations are addressed in Section \ref{sec:dec}, while conclusions are drawn in Section \ref{sec:conc}.

\section{Preliminaries} \label{sec:prel}
In this section, we describe the fundamental concepts necessary to familiarize with the structure, the encoding and the decoding of polar codes. 
This section can be skipped by readers having basic knowledge on the field, while we refer readers interested in further details and examples to \cite{polar_intro}.

\subsection{Channel polarization}
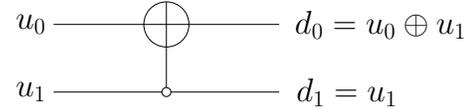
\begin{figure}[t]
  \centering
  \scalebox{0.6}{\input{figures/G_2}}
  \caption{Basic polarization kernel $\mathbf{G_2}$.}
  \label{figG2}
\end{figure}
The channel polarization phenomenon, introduced in \cite{arikan}, consists of a transformation which produces $N$ “synthetic” bit-channels from $N$ independent copies of a binary-input discrete memoryless channel (B-DMC). 
The new synthetic channels are then “polarized” in the sense that each of them can transmit a single bit at a different \emph{reliability}, i.e. with a different probability of being decoded correctly. 
If $N$ is large enough, the mutual information of the synthetic channels is either close to 0 (completely noisy channels) or close to 1 (perfectly noiseless channels), resulting in channels of extreme capacities.  
Polar codes provide a low-complexity scheme to construct polarized channels, where the fraction of noiseless channels tends to the capacity of the original B-DMC.  

The mathematical foundations of polar codes lie in the discovery of the polarization phenomenon of matrix $\mathbf{G_2} = \left[\begin{smallmatrix} 1&0\\1&1 \end{smallmatrix}\right]$, also known as \emph{basic polarization kernel}. 
This matrix enables the encoding of a two-bits input vector $\mathbf{u} = [u_0,u_1]$ into code word $\mathbf{d} = [d_0,d_1]$ as $\mathbf{d} = \mathbf{u} \cdot \mathbf{G_2}$, namely having $d_0 = u_0 \oplus u_1$ and $d_1 = u_1$. 
The polarization effect brought by $\mathbf{G_2}$ is more evident for binary erasure channels (BECs), where the transmitted bit is either received correctly or lost with probability $\delta$. 
If the input bits are decoded sequentially, bit $u_0$ cannot be recovered as $u_0 = d_0 \oplus d_1$ if any of the code bits has not been received, hence with probability $\delta_0 = (2 - \delta)\delta > \delta$. 
After $u_0$ has been correctly decoded, input bit $u_1$ can be decoded either as $u_1 = d_1$ or $u_1 = u_0 \oplus d_0$, hence it is sufficient that at least one of the two code bits has been received: consequently, failed decoding of this bit happens with probability $\delta_1 = \delta^2 < \delta$. 
As a result, input bit $u_0$ is transmitted over a degraded synthetic BEC with erasure probability $\delta_0 > \delta$, while $u_1$ is transmitted over an enhanced synthetic BEC with erasure probability $\delta_1 < \delta$. 

\subsection{Code design}
\begin{figure}
  \centering
  \includegraphics[width=0.45\textwidth]{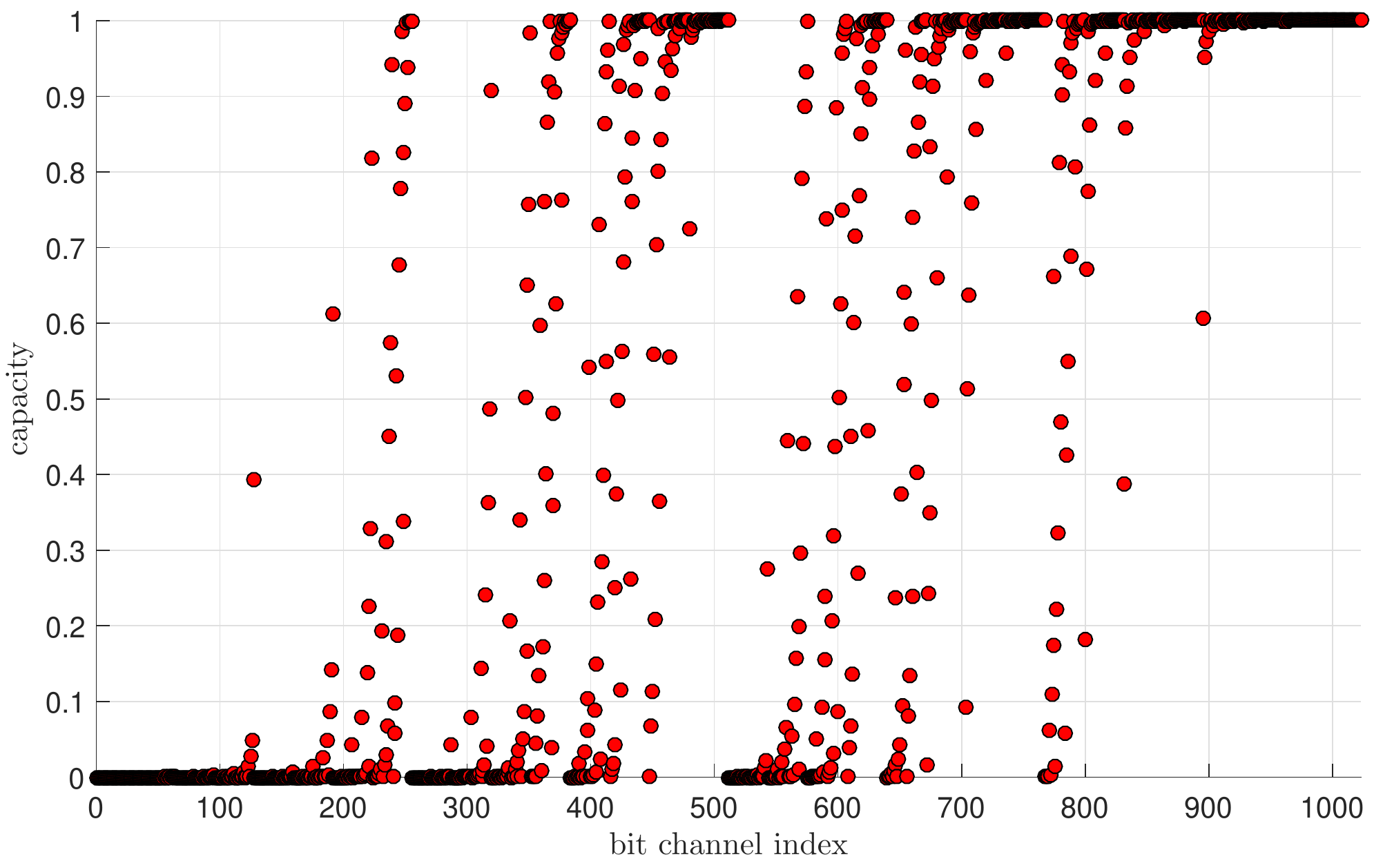}
  \caption{Virtual bit-channels capacities over a BEC $1/2$.}
  \label{fig:BEC_pol}
\end{figure}
Polar codes are based on the concatenation of several basic polarization kernels, creating a cascade reaction that speeds up the synthetic channels polarization while limiting encoding and decoding complexity. 
This concatenation generates a channel transformation matrix of the form $\mathbf{G_N} = \mathbf{G_2}^{\otimes n}$, defined by the $n$-fold Kronecker product of $\mathbf{G_2}$, and that can be recursively calculated as $\mathbf{G_N} = \left[\begin{smallmatrix} \mathbf{G_{N/2}}&0\\\mathbf{G_{N/2}}&\mathbf{G_{N/2}} \end{smallmatrix}\right]$.
While for $n \rightarrow \infty$ this construction creates channels that are either perfectly noiseless or completely noisy, for smaller values of $n$ the synthetic channels polarization may be incomplete, generating intermediary channels that are only partially noisy. 
Figure~\ref{fig:BEC_pol} shows the bit-channels polarization enabled by the concatenation of $n=10$ basic polarization kernels for a BEC with erasure probability $\delta = 1/2$. 

By construction, polar codes only allow for code lengths that are powers of two, in the form $N=2^n$; on the other hand the code dimension $K$, i.e. the number of information bits transmitted, can take any arbitrary value. 
The goal of code design of an $(N,K)$ polar code is to identify the $K$ best synthetic channels, namely the channels providing the highest reliability, and use them to transmit the information bits. 
The estimation of the reliability of each synthetic channel allows to sort them in reliability order and assign the $K$ information bits to the most reliable channels, whose indices constitute the \emph{information set} $\mathcal{I}$ of the code. 
The remaining $N-K$ indices form the \emph{frozen set} $\mathcal{F} = \mathcal{I}^C$ of the code, and the respective channels are “frozen”, not carrying any information. 

\subsection{Encoding}
\begin{figure}[t!]
  \centering
  \scalebox{0.37}{\input{figures/G_8}}
  \caption{Encoder of an $(8,4)$ polar code.}
  \label{figG8}
\end{figure}
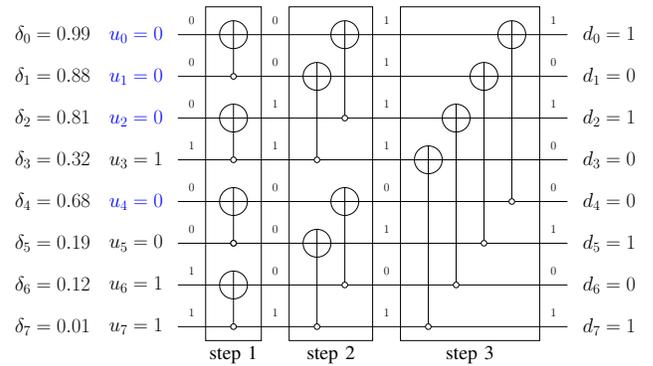
An $(N,K)$ polar code is defined by its channel transformation matrix $\mathbf{G_N} = \mathbf{G_2}^{\otimes n}$ and its information set $\mathcal{I}$. 
The generator matrix of the code is given by the sub-matrix of $\mathbf{G_N}$ composed by the rows whose indices are stored in $\mathcal{I}$. 
However, the recursive structure of $\mathbf{G_N}$ allows to reduce the encoding complexity with the introduction of an auxiliary input vector $\mathbf{u}$ of length $N$. 
This input vector $\mathbf{u} = [u_0,u_1,\ldots,u_{N-1}]$ is generated by assigning $u_i = 0$ if $i \in \mathcal{F}$, and storing the information bits in the remaining entries. 
Codeword $\mathbf{d} = [d_0,d_1,\ldots,d_{N-1}]$ is then calculated as 
\begin{equation}
\mathbf{d} = \mathbf{u} \cdot \mathbf{G_N}\ \text{.} \label{eq:polarGen}
\end{equation}
This matrix multiplication can be performed by processing multiple $\mathbf{G_2}$ matrix multiplications in parallel, reducing the encoding complexity to $\mathcal{O}(log_2(N))$. 
In fact, thanks to the recursive structure of $\mathbf{G_N}$, the decoding can be performed in $log_2(N)$ steps, each one composed by $N/2$ identical basic polarization kernels. 
Figure \ref{figG8} shows the structure of the encoder of a polar code of length $N=8$; moreover, it details the erasure probabilities of the synthetic BECs when the code is designed for a BEC of erasure probability $\delta = 1/2$, along with the frozen set $\mathcal{F} = \{ 0,1,2,4 \}$ obtained for a code of dimension $K=4$. 
This structure is used to encode message $\mathbf{c}=[1,0,1,1]$ in $n=3$ steps with the assistance of input vector $\mathbf{u} = [0,0,0,1,0,0,1,1]$ to obtain codeword $\mathbf{d} = [1,0,1,0,0,1,0,1]$.

\subsection{Decoding}
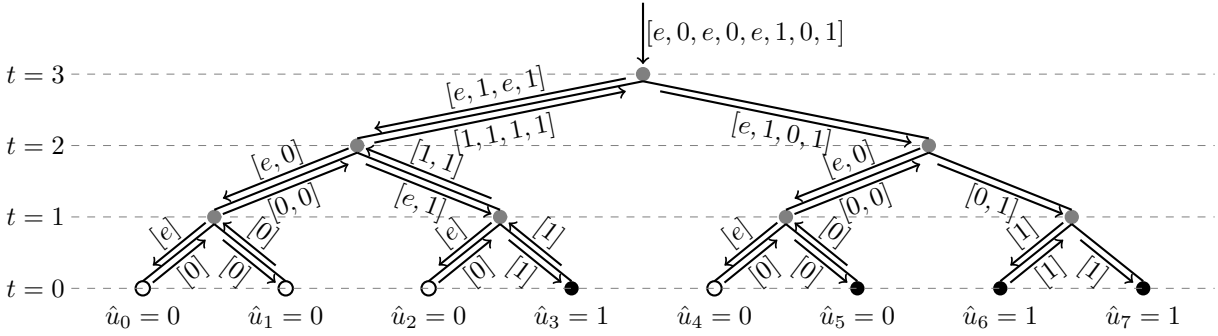
\begin{figure*}[t!]
  \centering
  \input{figures/sc-dec}
  \caption{SC decoding of an $(8,4)$ polar code over a BEC; white and black dots represent frozen and information bits.}
  \label{figSCDec}
\end{figure*}

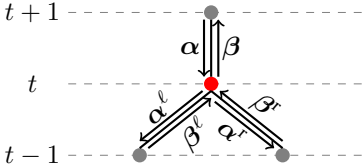
\begin{figure}[t!]
  \centering
  \input{figures/sc-node}
  \caption{SC decoding node.}
  \label{figSCNode}
\end{figure}

In \cite{arikan}, the decoding algorithm native to polar codes, called successive cancellation (SC), has been proposed as well. 
It can be represented as a depth-first binary tree search with priority to the left branch, where the leaf nodes are the $N$ bits to be estimated, and soft information on the received code bits is input at the root node, with a decoding complexity of $\mathcal{O}(N log_2(N))$. 
Figure~\ref{figSCDec} shows the decoding tree of the $(8,4)$ polar code depicted in Figure~\ref{figG8}, with black leaf nodes representing information bits and white ones frozen bits. 
Taking a node at stage $t$ as a reference, the message flow can be recursively described as shown in Figure~\ref{figSCNode}. 
The node uses the $2^t$ soft inputs $\bm{\alpha}$ received from its parent node to calculate $2^{t-1}$ soft outputs $\bm{\alpha^\ell}$ as $\alpha^\ell_i = f (\alpha_i,\alpha_{i+2^{t-1}})$ to be transmitted to its left child; later, it will combine the $2^{t-1}$ hard decisions $\bm{\beta^\ell}$ received from its left child with $\bm{\alpha}$ to calculate $2^{t-1}$ soft outputs $\bm{\alpha^r}$ for its right child as $\alpha^r_i = g(\alpha_i,\alpha_{i+2^{t-1}},\beta^\ell_i)$. 
Finally, the $2^{t-1}$ hard decisions $\bm{\beta^r}$ received from its right child are combined with $\bm{\beta^\ell}$ to calculate $2^t$ hard decisions $\bm{\beta}$ to be transmitted to its parent node as $\beta_i = \beta^\ell_i\oplus \beta^r_i$ if $i < 2^{t-1}$ and $\beta_i = \beta^r_{i-2^{t-1}}$ otherwise. 
When a leaf node is reached, the soft information is used to take a hard decision on the value of the information bits; frozen bits are always decoded as zeros. 

Update rules $f$ and $g$ for left and right child nodes respectively depend on the channel model. 
In the case of BEC, soft values belong to the set $\{0,1,e\}$, with $e$ representing an erasure, while hard decisions can only take values $0$ and $1$. 
Update rules are $f(\alpha_1,\alpha_2) = \alpha_1 \oplus \alpha_2$ and $g(\alpha_1,\alpha_2,\beta) = \alpha_2 \lor (\alpha_1 \oplus \beta)$; 
the erasure algebra imposes that $e \oplus \bullet = e$ and $e \lor \bullet = \bullet$. 
Continuing with the example of Figure~\ref{figG8}, Figure~\ref{figSCDec} shows the flux of soft and hard messages passed along the decoding tree if soft vector $\mathbf{y}=[e,0,e,0,e,1,0,1]$ has been received, allowing to correctly decode the message. 

\section{Advanced design} \label{sec:adv}
In this section, we introduce more advanced polar codeing principles. 
The discussion is not limited to frozen set calculation and encoding/decoding algorithms including techniques for rate matching and assistant bits design for 5G applications.  

\subsection{Frozen set design}
In general, the reliability order of the bit-channels depends on the channel condition and on the code length, and therefore is not universal. 
This non-universality poses huge practical problems in the construction of polar codes when a large range of code lengths and rates is envisaged. 
Many methods to design the frozen sets on-the-fly with limited complexity have been proposed \cite{comp_AWGN}. 
Along with Bhattacharyya parameter tracking, Ar{\i}kan initially proposed to use Monte-Carlo simulations to estimate bit-channel reliabilities \cite{arikan}. 
The density evolution (DE) method, initially proposed in \cite{DE_mori} and improved in \cite{tal}, can provide theoretical guarantees on the estimation accuracy, however at a high computational cost. 
A bit-channel reliability estimation method for AWGN channels based on Gaussian approximation (GA) of density evolution, termed as DE/GA method, has been proposed in \cite{trip_design}, giving accurate results with limited complexity. 
However, on-the-fly design increases coding latency too much to meet the 5G requirements. 

Recent studies on the partial reliability order imposed by the polarization effect on bit-channels \cite{part_ord,beta_exp,partHW} opened new opportunities in the generation of an universal reliability sequence to be used independently on channel conditions. 
These studies and intensive simulations lead the 5G standardization to propose a unique universal bit-channel reliability sequence to be used as a basis to extract the individual reliability sequence for each polar code considered in 5G.  
This sequence, composed by 1024 bit-channel indices sorted in reliability order, can be used regardless the channel conditions to design the frozen set of any polar code of length smaller or equal to 1024: it is in fact possible to extract sub-sequences for shorter codes directly from the universal reliability sequence, hence reducing the sequence storing space. 
This nested reliability structure represents a real breakthrough in polar code design, and it is probably the most lasting legacy of the standardization process. 
This impressive result has been achieved by taking into account also distance properties in short polar codes design \cite{mondelli,mk_dist_conf}, the use of list decoders and the presence of assistant bits in the code \cite{scal_exp}. 

\subsection{SC-based decoding}
Polar codes have been specifically conceived for SC decoding \cite{arikan}, with which they achieve channel capacity at infinite code length.  
Even if other decoding algorithms like BP \cite{BP_pc} and SCAN \cite{SCAN_pc} have been proposed, their poor error correction performance convinced 3GPP to focus the design envisaging an SC-based decoder at the receiver side. 
In the original formulation of SC, soft information was described in terms of likelihoods \cite{arikan}, that are numerically unstable and not suitable for hardware implementations. 
This instability has been reduced using log-likelihoods first \cite{balatsoukas_SCL_HW}, and finally eliminated with log-likelihood ratios (LLRs) \cite{balatsoukas}. 
The SC algorithm can be implemented in both software and hardware with low complexity \cite{leroux,Gal_SW}, however its error-correction performance is mediocre at practical code lengths. 
Thus, many attempts have been made to overcome this issue \cite{SCFlip14,SCF-WCNC18}. 

Eventually, a list-based decoding approach to polar codes (SCL) was introduced in \cite{tal_list}. 
The idea is to let a group of SC decoders work in parallel, maintaining different codeword candidates, or paths, at the same time. 
Every time a leaf node is reached, the bit is estimated as both 1 and 0, doubling the number of codeword candidates; a path metric for each candidate is then calculated to discard less likely candidates, hence limiting the number of paths. 
SCL substantially improves the error-correction performance of SC at moderate code lengths, especially when the code is concatenated to an outer code acting as a genie, e.g. a cyclic redundancy check (CRC), at the cost of an augmented complexity. 
CRC-assisted (CA) SCL has been used as a baseline in 5G error-correction performance evaluations, in particular with a list size equal to 8 \cite{RGPP_AH1}. 

While the effectiveness of SCL improves as the list size increases, its implementation complexity increases as well. 
To limit the rise in complexity, various approaches have been proposed for software and hardware decoders alike. 
Partitioned SCL \cite{PSCL-ICASSP} and its evolutions \cite{PSCL-GLOBECOM,PSCL-JETCAS} consider different list sizes at different stages of the SC tree, reducing the memory requirements at higher stages. 
SC-Stack decoding \cite{Stack:Original} expands only the most probable candidate thanks to a priority queue. 
Adaptive SCL decoding \cite{Adaptive} foresees increasing list sizes in case of failed decoding, while a hardware decoder with flexible list size has been proposed in \cite{BlindDetHW}.

\subsection{Rate matching}
By definition, the polar code length $N$ is limited to powers of two, while any number $K$ of bit-channels can be included in the information set. 
This code length constraint represents a limitation for typical 5G applications, where the amount of information $A$ is fixed and a codeword of length $E$ is needed to achieve the desired rate $R=A/E$.
Rate matching for polar codes becomes thus a length matching problem, and has been solved for 5G through classical coding theory techniques as puncturing, shortening and extending \cite{RiU08}. 

Both puncturing and shortening reduce the length of a mother code by not transmitting code bits in a predetermined pattern, called \emph{matching pattern}; the difference lies in the meaning of the code bits belonging to the matching pattern. 
In puncturing, the untransmitted code bits are treated as erased at the decoder, while shortening introduces a sub-code imposing the untransmitted code bits to assume a fixed value, typically zero, such that their value is already known by the decoder. 
It has been observed that for polar codes, shortening works better for high rates, and puncturing for low rates \cite{punct_paper}. 

Puncturing $U$ code bits deteriorates the bit-channel reliabilities, additionally causing $U$ bit channels, termed as \emph{incapable bits}, to be completely unreliable; the position of these bits can be calculated on the basis of the matching pattern \cite{short_mat}.  
In order to avoid catastrophic error-correction performance degradation under SC decoding, incapable bits must be frozen to prevent random decisions at the decoder. 
Shortening, on the contrary, improves the bit channel reliabilities by introducing overcapable bits, i.e. bits with conceptually infinite reliability that are always correctly decoded \cite{punct_paper}. 
However, code bits in the matching pattern must depend on frozen bits only, which forces to freeze the most reliable bit-channels worsening the error correction capabilities of the code.

Rate matching alters the code reliability sequence so that bit-channel reliabilities need to be recalculated to find the optimal frozen set; 
DE/GA method is used to evaluate new bit-channel reliabilities in \cite{wang_liu} for shortening and in \cite{short_mat,Chandesris_ICC17} for puncturing. 
The latency increase due to on-the-fly reliability calculation brought 3GPP to prefer an alternative approach, namely to design the matching pattern on the basis of the frozen set \cite{isit_punc,ARQ_punct}. 
This significantly reduces the code design complexity, albeit at the cost of an increased block error rate (BLER). 
The joint optimization of frozen set and matching pattern, as proposed in \cite{short_milo}, has been evaluated too, however it has been discarded due to its excessive complexity. 

When the code length $E$ is slightly larger than a power of two, a large number of bits would not be transmitted if puncturing or shortening were to be applied, since the mother polar code length would have to be selected as the successive power of two. 
In this case, extending a smaller polar code, i.e. retransmitting some code bits, may be preferable in terms of decoding latency and error correction performance \cite{ARQ_polar} since a smaller mother polar code decoder can be used combining the likelihoods of retransmitted bits. 
The presence of repeated code bits due to retransmission alters the bit-channels reliability, however without giving restrictions to code design as for puncturing or shortening \cite{ARQ_punct}. 
This allowed 3GPP to focus on the design of puncturing and shortening mechanism while leaving the repetition scheme as simple as possible. 
As a result, a circular buffer rate matching \cite{circ_buff_polar} scheme has been adopted incorporating all three techniques in an elegant and simple way. 

\begin{figure*}[t!]
\centering
\input{figures/5Gchain.tikz}
\caption{5G polar codes encoding chain; yellow, red and orange blocks are implemented in downlink, uplink and both respectively. }
\label{fig:5Gchain}
\end{figure*}
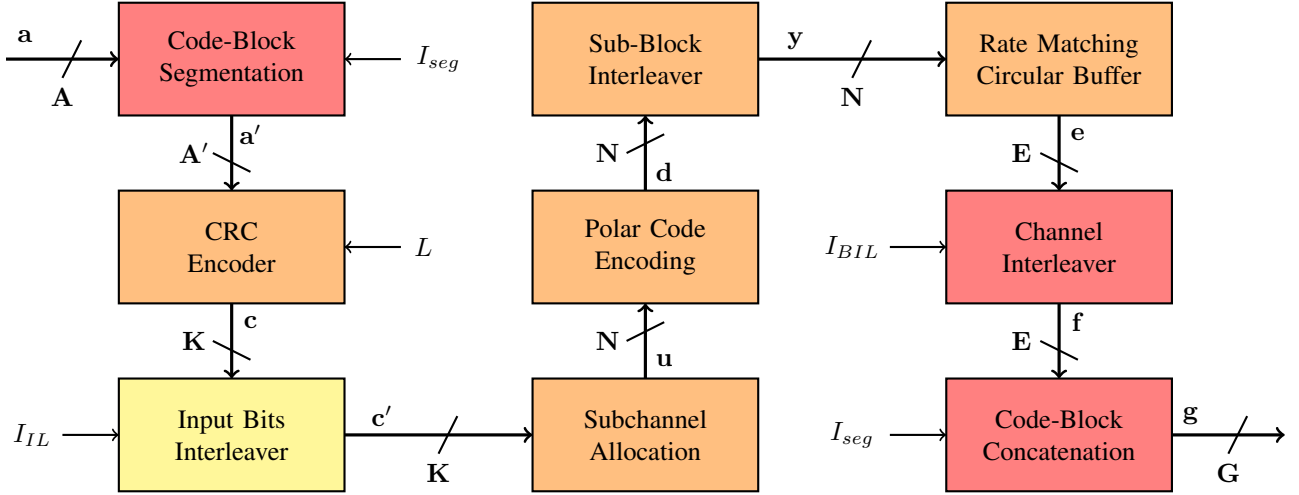

\subsection{Assistant bits design}
Assistant bits can be broadly defined as additional bits that help the decoding of the polar code, usually improving its error-correction performance. 
Their introduction dates back to the conception of SCL decoding, for which it has been noticed that the introduction of an outer CRC code improved the error-correction performance of the code \cite{tal_list,CRC_aid}. 
In general, it has been proven that the minimum distance of polar codes can be dramaticaly improved by adding an outer code \cite{PC_conc}, and SCL decoders can fully exploit this improved code spectrum \cite{PC_const_list}. 
Parity-checks (PC) have emerged as an alternative to CRC due to their simplicity and flexibility \cite{PC_conc,dynFroz}. 
Introducing PC bits in the middle of the decoding instead of a single CRC check at the end makes it easier to tune the polar code spectrum; bit-channels corresponding to minimum weight rows are the best candidates to host parity checks \cite{PC_min_ham}. 
3GPP decided to adopt a low complexity PC design based on shift registers \cite{PC_hua} in conjunction with a classical CRC, as proposed in \cite{PC_CRC}, to improve the code design flexibility.

The insertion of an interleaver between the CRC encoder and the polar encoder may have a positive impact on the performance of the code due to the induced change in the number of minimum weight codewords \cite{conc_polar_cyc}.
The interleaver is used to turn the CRC into a distributed CRC, in that a CRC remainder bit is assigned to a bit-channel as soon as all the information bits involved in its parity check have been assigned as well. 
This feature is used in 5G polar codes to reduce the decoding complexity by early terminating the decoding if an incorrect check is met, providing an additional false alarm rate (FAR) mitigation mechanism \cite{dist_CRC_aid}. 

\subsection{Polar coded modulation}
If low-order modulation schemes like quadrature phase shift keying (QPSK) are used, the bit-channels polarization is not affected since all modulated symbols have uniform reliability. 
When larger constellations are used, as envisaged by 5G standard, polar-coded modulation (PCM) \cite{PCM} can be used to fully exploit the polarization effect at higher-order modulation \cite{eff_PCM}. 
However, canonical PCM requires an additional polarization matrix whose size depends on the modulation scheme used \cite{64QAM}, resulting in an increased decoding latency \cite{PC_CM}. 

A channel interleaver has been introduced by 3GPP as a low-complexity alternative to PCM; this technique, termed as bit-interleaved polar-coded modulation (BIPCM) \cite{BIPCM_conf}, improves the diversity gain under high-order modulation without increasing the code complexity \cite{eff_BIPCM}. 
In BIPCM, the channel is considered as a set of parallel bit channels which can be combined with polar coded bits. 
Carefully mapping coded bits to modulation symbols offers a certain gain over the conventional random interleaving \cite{2mary_PCM}, which can be further improved designing interleavers adaptive to channel selectivity \cite{PDM_OFDM}. 
Finally, the correlation among coded bits mapped into the same symbol allows to combine the demapping and deinterleaving units with the SC decoder to perform the decoding directly on the LLRs of the received symbols \cite{SC_BIPCM}. 

\section{Polar Code Encoding in 5G} \label{sec:5G}
\begin{figure*}
\begin{subfigure}{.5\textwidth}
  \centering
  \includegraphics[width=0.95\textwidth]{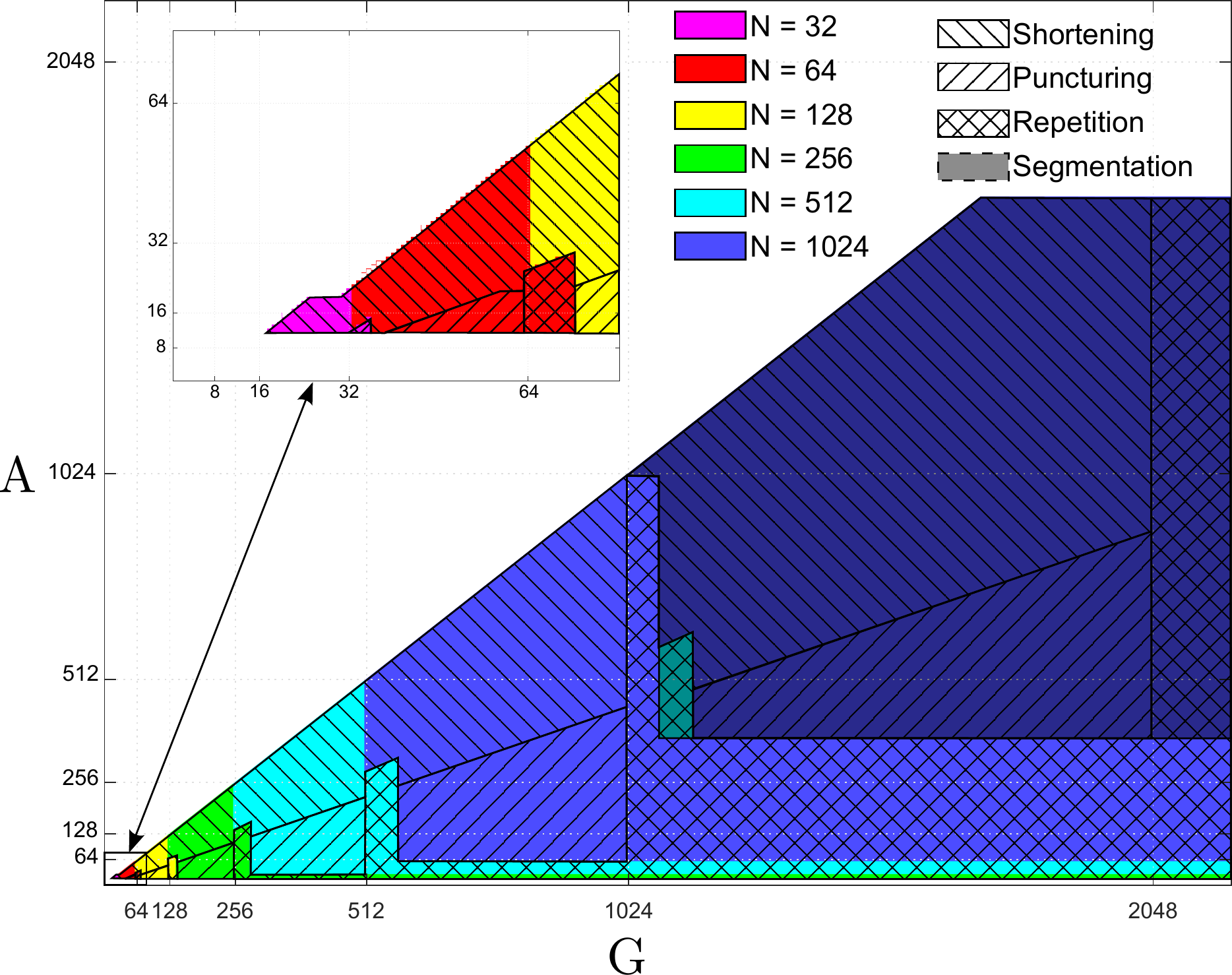}
  \caption{PUCCH/PUSCH}
\end{subfigure}
\begin{subfigure}{.5\textwidth}
  \centering
  \includegraphics[width=0.95\textwidth]{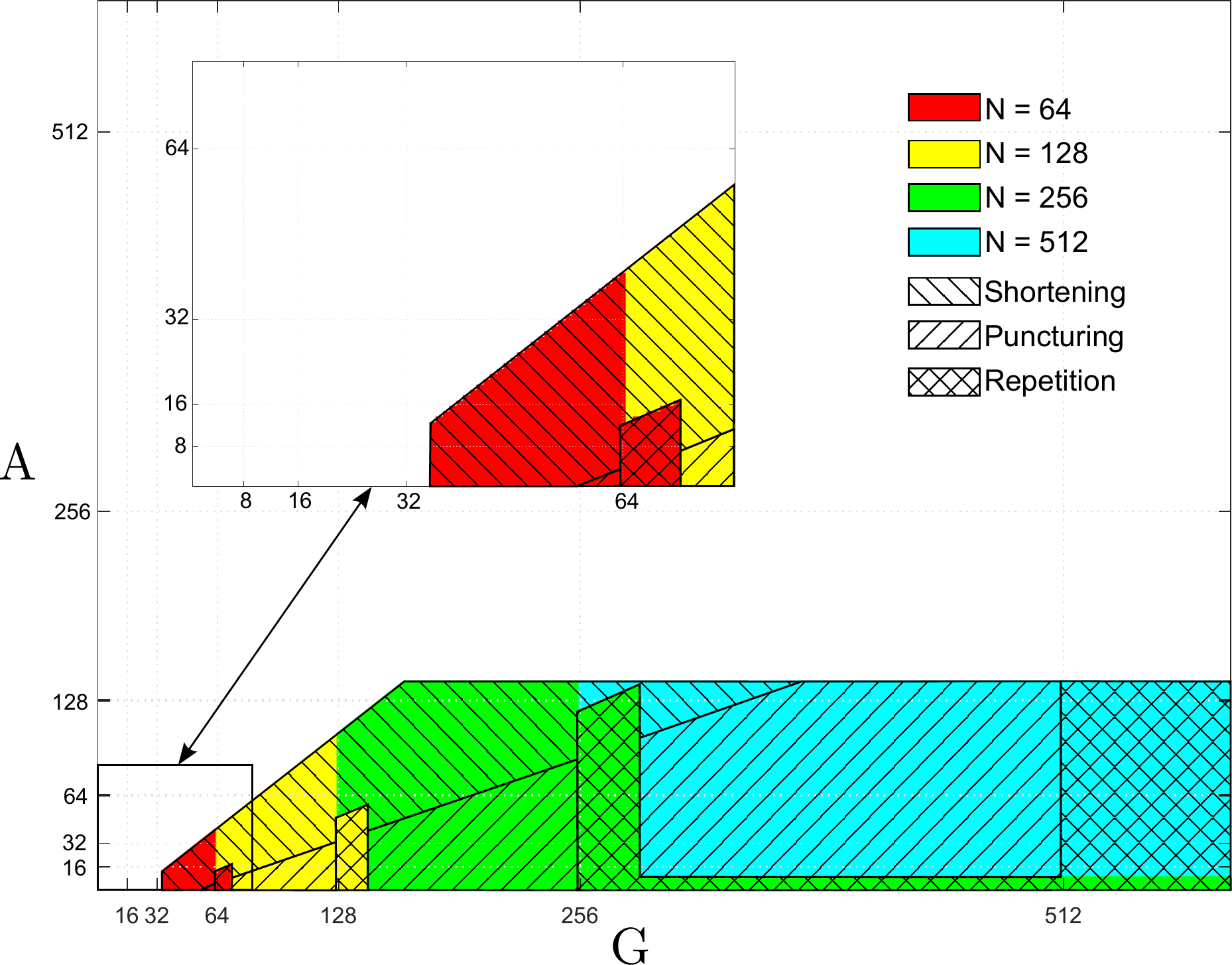}
  \caption{PDCCH/PBCH}
\end{subfigure}
\caption{Mother polar code length $N$ and rate matching scheme for admitted values of message length $A$ and payload length $G$. }
\label{fig:U_D_leng}
\end{figure*}

In this section we describe in detail the framework adopted for the encoding of polar codes in 5G standard. 
In the following, the notation introduced in the 3GPP technical specification \cite{3GPP_R15} will be used. 
Polar codes are used to encode the uplink control information (UCI) over the physical uplink control channel (PUCCH) and the physical uplink shared channel (PUSCH). 
In the downlink, polar codes are used to encode the downlink control information (DCI) over the physical downlink control channel (PDCCH), and the payload in the physical broadcast channel (PBCH). 
Table \ref{tab:sched_chan} summarizes the encoding chain parameters and bounds. 

In 5G applications, the number of information bits, $A$, is fixed and a codeword of length $E$ is created to achieve the desired rate $R=A/E$ required by upper communication layers. 
To accommodate polar codes to this requirement, a mother polar code of length $N = 2^n$ is initially constructed, and the desired code length $E$ is matched via puncturing, shortening or repetition. 
The mother code length $N$ is lower bounded by $N_{min} = 32$, while the value of the upper bound $N_{max}$ depends on the channel used, being $N_{max} = 512$ for downlink and $N_{max} = 1024$ for uplink. 
An ulterior upper bound is imposed by the minimal accepted code rate of $\frac{1}{8}$. 

Figure~\ref{fig:5Gchain} portrays the set of encoding operations envisaged by the 5G polar codes framework. 
Vector $\mathbf{a}$ contains the $A$ information bits to be transmitted using a payload of $G$ code bits. 
Depending on code parameters, the message may be split and segmented into two parts, which are encoded separately and transmitted together. 
Each segmented vector $\mathbf{a'}$ of length $A'$ will be encoded into an $E$-bit polar codeword. 
To every $A'$-bit vector, an $L$-bit CRC is attached. 
The resulting vector $\mathbf{c}$, constituted of $K=A'+L$ bits, is passed through an interleaver. 
On the basis of the desired code rate $R$ and codeword length $E$, a mother polar code of length $N$ is designed, along with the relative bit channel reliability sequence and frozen set. 
The interleaved vector $\mathbf{c'}$ is assigned to the information set along with possible parity-check bits, while the remaining bits in the $N$-bit $\mathbf{u}$ vector are frozen. 
Vector $\mathbf{u}$ is then encoded as $\mathbf{d=u \cdot G_N}$, where $\mathbf{G_N} = \mathbf{G_2}^{\otimes n}$ is the generator matrix for the selected mother code. 
After encoding, a sub-block interleaver divides $\mathbf{d}$ into $32$ equal-length blocks, scrambles them and creates $\mathbf{y}$, that is fed into the circular buffer. 
For rate matching, puncturing, shortening or repetition are applied to change the $N$-bit vector $\mathbf{y}$ into the $E$-bit vector $\mathbf{e}$. 
A channel interleaver is finally applied to compute the vector $\mathbf{f}$, that is now ready to be modulated and transmitted as $\mathbf{g}$ after concatenation, if required. 

Parameters $A$ and $E$ are bounded according to the channel used, obviously having $A \leq E$. 
In the uplink, $12 \leq A \leq 1706$, while for $A\le11$ different block codes are used. 
The agreed codeword length range is $18 \leq E \leq 8192$, however the payload length $G$ can be larger since $G \leq 16384$: in this case, the information bits may be divided into two polar codewords through segmentation. 
In the downlink, $A$ is upper bounded by 140 for PDCCH, however in this case if $A \le 11$ the message is padded with zeros to reach $A = 12$. 
Due to the presence of the CRC, $E$ is lower bounded by 25, while $E \leq 8192$ as for uplink. 
For PBCH, only one code is accepted with parameters $A = 32$ and $E = 864$. 
The flags $I_{IL}$ and $I_{BIL}$ refer to the activation of the input bits interleaver and the channel interleaver, respectively. 
The number of the two types of assistant PC bits are given by $n_{PC}$ and $n_{PC}^{wm}$ (see Section \ref{subsec:subch} for details). 
Table \ref{tab:sched_chan} summarizes the encoding chain parameters, while Figure~\ref{fig:U_D_leng} shows with a line pattern the rate matching scheme and mother polar code length used for different combinations of payload length $G$ and number of information bits $A$ for both uplink and downlink channels; segmentation for PUCCH is identified by area darkening. 

\subsection{Message segmentation} 
\label{subsec:seg}
Given message length $A$ and payload length $G$, the information may be decomposed in two blocks and encoded separately. 
Segmentation is activated by flag $I_{seg}$, and in particular it may be activated for PUCCH and PUSCH UCIs ($I_{seg} = 1$) while it is always bypassed for PBCH payloads and PDCCH DCIs ($I_{seg} = 0$). 
Segmentation is necessary for the uplink when message or payload lengths are longer than maximum mother polar code length and shortening or puncturing is used as rate matching mechanism; this is not necessary for the downlink since the message length is always shorter than mother polar code length and repetition is used for large payload sizes. 
When code parameters satisfy the condition $(A \geq 1013) \vee (A \geq 360 \wedge G \geq 1088)$, segmentation is activated.
In this case, the message is divided into two parts of length $A' = \lceil A/2 \rceil$; if $A$ is odd, the first message is composed by the first  $\lfloor A/2 \rfloor$ bits with the addition of a zero padding at the beginning. 
The code length, usually set to $E=G$, has to be changed accordingly, namely setting $E = \lceil G/2 \rceil$. 

\subsection{Mother polar code length and rate matching selection} \label{subsec:param}

\begin{table*}[t!]
\begin{center}
\caption{Code parameters and bounds for different channels gathered from \cite{3GPP_R15}.}
\label{tab:sched_chan}
\begin{tabular}{|l|c|c|c|c|c|c|c|}
 \cline{3-8}
 \multicolumn{2}{c|}{} & \multicolumn{4}{c|}{Uplink} & \multicolumn{2}{c|}{Downlink} \\
 \cline{3-8}
 \multicolumn{2}{c|}{} & \multicolumn{4}{c|}{PUCCH/PUSCH} & \multirow{4}{*}{PDCCH} & \multirow{4}{*}{PBCH} \\
 \cline{3-6}
 \multicolumn{2}{c|}{} & \multicolumn{2}{c|}{$A \geq 20$} & \multicolumn{2}{c|}{$12 \leq A \leq 19$} &  &\\
 \cline{3-6}
 \multicolumn{2}{c|}{} & $(A \geq 1013) \vee$ & $(A < 360) \vee$ & \multirow{2}{*}{$E-A \leq 175$} & \multirow{2}{*}{$E-A > 175$} &  & \\
 \multicolumn{2}{c|}{} & $(A \geq 360 \wedge G \geq 1088)$ & $(A < 1013 \wedge G < 1088)$ &  & & & \\
 \hline
 Max polar code exponent & $n_{max}$ & \multicolumn{4}{c|}{10} & \multicolumn{2}{c|}{9} \\
 \hline 
 Input bits inter. flag & $I_{IL}$ & \multicolumn{4}{c|}{0} & \multicolumn{2}{c|}{1} \\
 \hline 
 Channel inter. flag & $I_{BIL}$ & \multicolumn{4}{c|}{ 1 } & \multicolumn{2}{c|}{0} \\
 \hline 
 Segmentation flag & $I_{seg}$ & 1 & \multicolumn{3}{c|}{0} & \multicolumn{2}{c|}{0} \\
 \hline 
 Max payload length & $G_{max}$ & 16384 & \multicolumn{3}{c|}{8192} & 8192 & 864 \\
 \hline 
 Min payload length & $G_{min}$ & \multicolumn{2}{c|}{31} & \multicolumn{2}{c|}{18} & 25 & 864 \\
 \hline
 Max message length & $A_{max}$ & \multicolumn{4}{c|}{1706} & 140 & 32 \\
 \hline
 Min message length & $A_{min}$ & \multicolumn{4}{c|}{12} & 1 & 32 \\
 \hline
 CRC length & $L$ & \multicolumn{2}{c|}{11} & \multicolumn{2}{c|}{6} & \multicolumn{2}{c|}{24} \\
 \hline 
 \# of PC bits & $n_{PC}$ & \multicolumn{2}{c|}{0} & \multicolumn{2}{c|}{3} & \multicolumn{2}{c|}{0} \\
 \hline 
 \# of row-weight PC bits & $n_{PC}^{wm}$ & \multicolumn{2}{c|}{0} & 0 & 1 & \multicolumn{2}{c|}{0} \\
 \hline 
\end{tabular}
\end{center}
\end{table*}

The mother polar code length $N = 2^n$ is a crucial parameter in the encoding process. 
Its logarithm $n$ is selected as
\begin{equation}
n = \max(\min(n_1,n_2,n_{max}),n_{min})~,
\end{equation}
where $n_{min}$ and $n_{max}$ give a lower and an upper bound on the mother code length, respectively. In particular, $n_{min} = 5$, while $n_{max} = 9$ for the downlink control channel, and $n_{max} = 10$ for the uplink. Parameter $n_2$ gives an upper bound on the code based on the minimum code rate admitted by the encoder, i.e. $\frac{1}{8}$; as a consequence, $n_2 = \lceil \log_2(8K) \rceil$. 
Finally, the value of $n_1$ is bound to the selection of the rate-matching scheme. It is in fact usually calculated as $n_1 = \lceil \log_2(E) \rceil$, so that $2^{n_1}$ is the smallest power of two larger than $E$. 
However, a correction factor is introduced to avoid a too severe rate matching: if\footnote{$\{ x \} = x - \lfloor x \rfloor$ represents the fractional part of a real number $x$.} $\{ \log_2(E) \} < 0.17$, i.e. if the smallest power of two larger than $E$ is too far from $E$, the parameter is set to $n_1 = \lfloor \log_2(E) \rfloor$, and an additional constraint on the code dimension is added, namely $K < \frac{9}{16} E$, to assure that $K < N$.

If a mother polar code of length $N>E$ is selected, the mother polar code will be punctured or shortened, depending on the code rate, before the transmission. 
In particular, if $\frac{K}{E} \leq \frac{7}{16}$, the code will be punctured, otherwise it will be shortened. 
On the contrary, if $N<E$, repetition is applied and some encoded bits will be transmitted twice; in this case, the code construction assures that $K<N$. 

\subsection{CRC encoding} \label{subsec:CRC}
A CRC of $L$ bits is appended to the $A'$ message bits stored in $\mathbf{a'}$, resulting in a vector $\mathbf{c}$ of $A'+L$ bits. 
The possible CRC generator polynomials are the following:
\begin{align*}
g_{6}(x)  =& x^6+x^5+1 \\
g_{11}(x) =& x^{11}+x^{10}+x^9+x^5+1 \\
g_{24}(x) =& x^{24}+x^{23}+x^{21}+x^{20}+x^{17}+x^{15}+x^{13}+x^{12}+ \\
        & +x^{8}+x^{4}+x^{2}+x+1
\end{align*}
The polynomial $g_{24}(x)$ is used for the payload in PBCH and DCIs in the PDCCH, where a larger number
of assistant bits are necessary to enable early termination in the case of failures. 
Polynomials $g_{6}(x)$ and $g_{11}(x)$ are used for UCIs, in the case $12\le A\le 19$ and $A\ge 20$, respectively.
The CRC shift register is initialized by all zeros for UCIs and for the PBCH payloads, and to all ones for the DCIs.
Moreover, for DCIs, the CRC parity bits are ``scrambled" according to a radio network temporary identifier (RNTI) $x^{rnti}_0,x^{rnti}_1,...,x^{rnti}_{15}$, i.e. the RNTI is masked in the last 16 CRC bits calculated by $g_{24}(x)$ as $c_{A+8+k} = c_{A+8+k} \oplus x^{rnti}_{k}$ for $k = 0,\dots,15$. 

\subsection{Input bits Interleaver} \label{subsec:ILV1}
\begin{figure}
  \centering
  \includegraphics[width=0.45\textwidth]{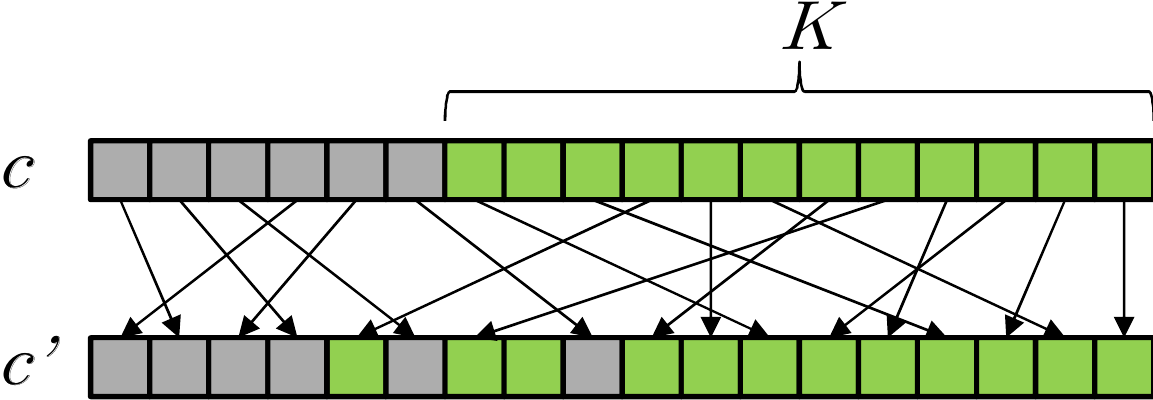}
  \caption{Input bits interleaver design.}
  \label{fig:IB_inter}
\end{figure}

The $K$ bits obtained from the CRC encoder are interleaved before being inserted into the information set of the mother polar code. 
This feature can be used to reduce the decoding complexity by early terminating the decoding if an incorrect check is met, providing an additional false alarm rate (FAR) mitigation.
The interleaver is selected through the flag parameter $I_{IL}$, being activated for PBCH payloads and PDCCH DCIs ($I_{IL}=1$) and bypassed in the case of PUCCH and PUSCH UCIs ($I_{IL}=0$); early decoding termination has been in fact considered crucial for downlink, to reduce the computation effort of mobile devices, while higher computation capabilities of base stations made it a less attracting feature for uplink. 

By construction, the number of interleaved bits is upper bounded by $K_{IL}^{max} = 164$. 
The $K$ bits from the previous step are padded to reach the length of 164 bits and permuted according to the sequence $\Pi_{IL}^{max}$ presented in Table~\ref{tab:IB_inter}. 
It is worth noticing that last bits are not interleaved to maintain the majority of CRC bits at the end of the message.  
In more detail, the parameter $h = K_{IL}^{max} - K$ is calculated. 
Starting from the entry at index $0$, all elements of $\Pi_{IL}^{max}$ are compared to $h$, and in the case that they are larger, they are stored in $\Pi$. 
Finally, $h$ is subtracted from all the entries of $\Pi$, such that $\Pi$ contains all the positive integers smaller than $K$ in permuted order. 
The interleaving function is applied to $\mathbf{c}$, and the $K$-bit vector $\mathbf{c'} = \{ c_{\Pi(0)},\dots,c_{\Pi(K-1)} \}$ is obtained. 
This interleaver enables early termination since every CRC remainder bit is placed after its relevant information bits. 
Its functioning is depicted in Figure~\ref{fig:IB_inter}: the $K$ information bits are stored at the end of the fixed scrambler mixing them, while $\mathbf{c'}$ is obtained extracting the green squares from left to right.
\begin{table}
\begin{center}
\caption{Input bits interleaver pattern sequence $\Pi_{IL}^{max}$ (bold integers represent CRC bit indices).}
\label{tab:IB_inter}
\resizebox{0.48\textwidth}{!}{
\setlength{\extrarowheight}{1.7pt}
\begin{tabular}{|cccccccccccc|}
 \hline
  0 &   2 &   4 &   7 &   9 &  14 &  19 &  20 &  24 &  25 &  26 &  28 \\
 31 &  34 &  42 &  45 &  49 &  50 &  51 &  53 &  54 &  56 &  58 &  59 \\
 61 &  62 &  65 &  66 &  67 &  69 &  70 &  71 &  72 &  76 &  77 &  81 \\
 82 &  83 &  87 &  88 &  89 &  91 &  93 &  95 &  98 & 101 & 104 & 106 \\
108 & 110 & 111 & 113 & 115 & 118 & 119 & 120 & 122 & 123 & 126 & 127 \\
129 & 132 & 134 & 138 & 139 & \textbf{140} &   1 &   3 &   5 &   8 &  10 &  15 \\
 21 &  27 &  29 &  32 &  35 &  43 &  46 &  52 &  55 &  57 &  60 &  63 \\
 68 &  73 &  78 &  84 &  90 &  92 &  94 &  96 &  99 & 102 & 105 & 107 \\
109 & 112 & 114 & 116 & 121 & 124 & 128 & 130 & 133 & 135 & \textbf{141} &   6 \\
 11 &  16 &  22 &  30 &  33 &  36 &  44 &  47 &  64 &  74 &  79 &  85 \\
 97 & 100 & 103 & 117 & 125 & 131 & 136 & \textbf{142} &  12 &  17 &  23 &  37 \\
 48 &  75 &  80 &  86 & 137 & \textbf{143} &  13 &  18 &  38 & \textbf{144} &  39 & \textbf{145} \\
 40 & \textbf{146} &  41 & \textbf{147} & \textbf{148} & \textbf{149} & \textbf{150} & \textbf{151} & \textbf{152} & \textbf{153} & \textbf{154} & \textbf{155} \\
\textbf{156} & \textbf{157} & \textbf{158} & \textbf{159} & \textbf{160} & \textbf{161} & \textbf{162} & \textbf{163}  & &  &  &  \\
\hline
\end{tabular}
}
\end{center}
\end{table}

\subsection{Subchannel allocation and PC bits calculation} \label{subsec:subch}
At this step, vector $\mathbf{c'}$ is expanded in the $N$-bit input vector $\mathbf{u}$ with the addition of assistant and frozen bits.
To begin with, $n_{PC}$ parity-check bits are inserted within the $K$ information and CRC bits. 
The mother polar code is hence a $(N,K')$ code, with $K' = K + n_{PC}$. 
To create the input vector $\mathbf{u}$ to be encoded, the frozen set of subchannels needs to be identified. 
The number and position of frozen bits depend on $N$, $E$, and the selected rate-matching scheme.
Initially, the frozen set $\bar{Q}_F^N$ and its complementary set, the information set $\bar{Q}_I^N$, are computed based on the universal reliability sequence $Q_0^{N_{max}-1}$ \cite{3gpp_polarSequence} and the rate matching strategy. 
Later, information bits are assigned to $\mathbf{u}$ according to the information set. 
Finally, assistant PC bits are calculated and stored in $\mathbf{u}$, if necessary. 

\begin{figure}
  \centering
  \includegraphics[width=0.45\textwidth]{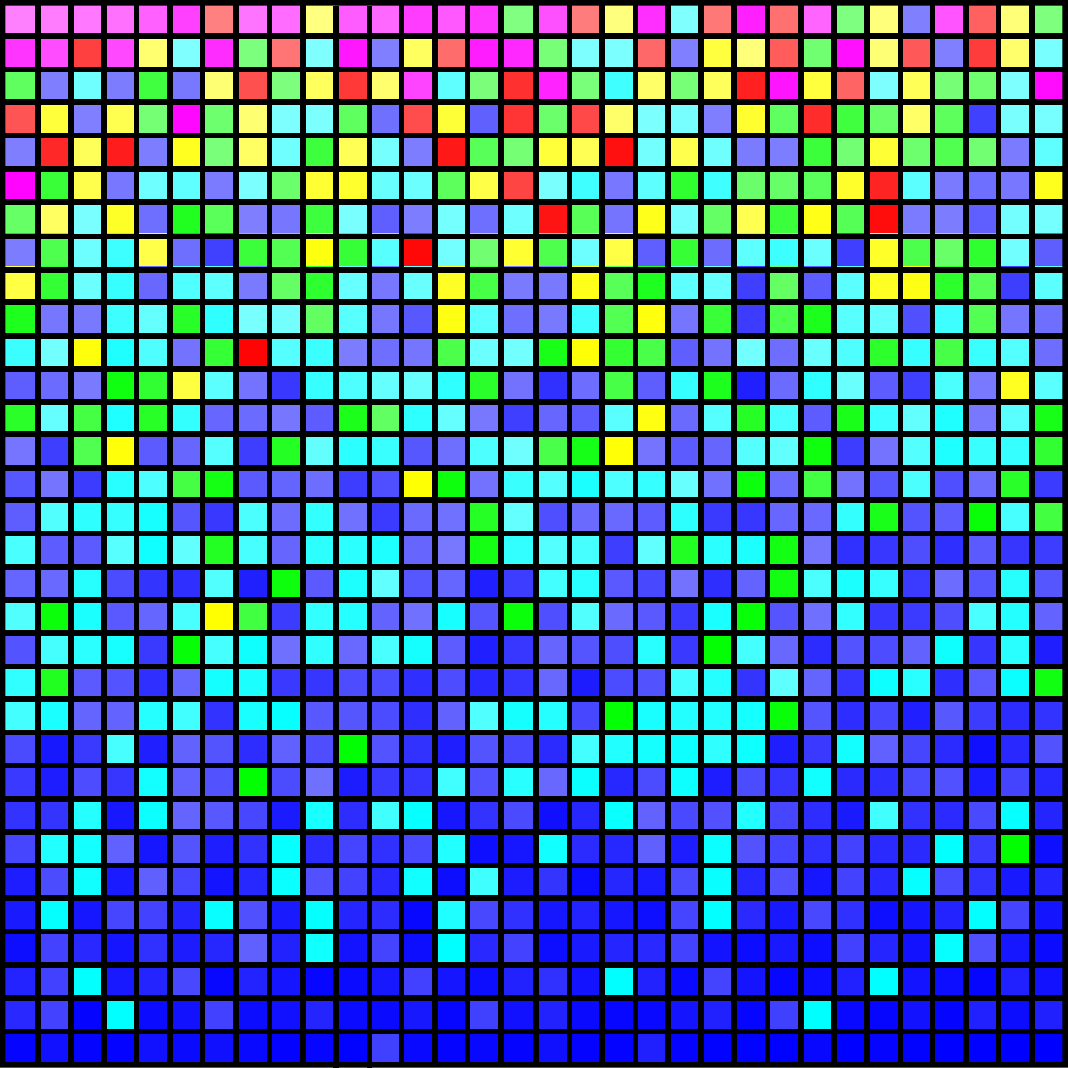}
  \caption{Universal reliability sequence $Q_0^{N_{max}-1}$. Magenta, red, yellow, green, cyan and blue squares represent entries smaller than 32, 64, 128, 256, 512 and 1024 respectively. Brightness indicates reliability in the color interval. }
  \label{fig:rel_seq}
\end{figure}

The universal reliability sequence $Q_0^{N_{max}-1}$ is the list of all positive integers smaller than $1024$ sorted in reliability order, from the least reliable to the most reliable; indices smaller than $N$ are extracted neatly in the creation of the frozen set of a mother polar code of length $N$. 
The full sequence can be found in \cite{3GPP_R15} and is depicted in Figure~\ref{fig:rel_seq} to illustrate the bit-channels distribution. 
The $1024$ squares represent all the bit-channels row-by-row from the least reliable in the top-left corner to the most reliable in the bottom-right corner. 
There are 32 magenta subchannels, that are relative to the bit indices 0 to 31, and 32 red subchannels, referring to bit indices 32 to 63. 
The 64 yellow ones are relative to bit indices 64 to 127, the 128 green ones to bits 128 to 255, the 256 azure ones to bits 256 to 511, and the 512 blue ones to bits 512 to 1023. 
Darker shades of each color represent larger indices, while lighter shades are smaller ones. 
As an example, if $N=512$, the red, yellow, green and azure entries of the sequence are extracted maintaining their relative order, for a total of 512 ordered indices. 
In the case that $N=256$ is selected, only the red, yellow and green ones are extracted, and so on.

\subsubsection{Frozen set $\bar{Q}_F^N$}
\begin{figure*}
  \centering
  \includegraphics[width=0.95\textwidth]{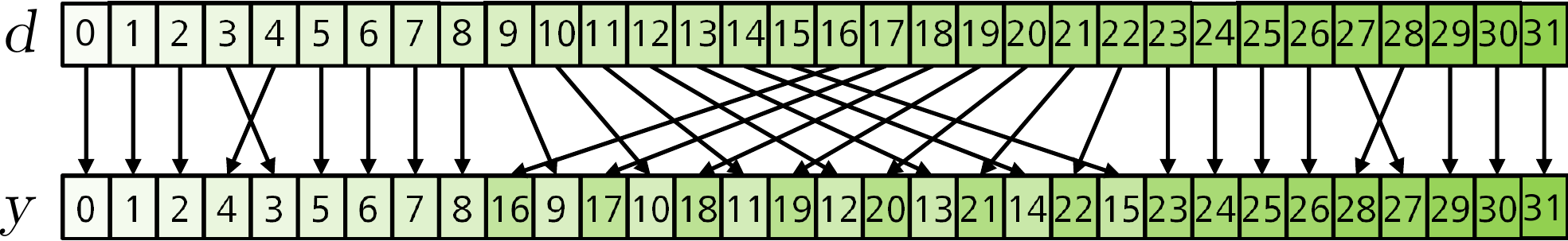}
  \caption{Design of the sub-block interleaver $J$.}
  \label{fig:SB_inter}
\end{figure*}
The first bits identified in the frozen set correspond to the indices of the $U = N - E$ untransmitted bits, i.e. the bits eliminated from the codeword by the rate-matching scheme. 
These indices correspond to the first $U$ or the last $U$ codeword bits in the case of puncturing and shortening, respectively, as explained in Section~\ref{subsec:RM}. 
Due to the presence of sub-block interleaver $J$, the actual indices to be included in the frozen set correspond to the first or the last after interleaving; details on this sub-block interleaver can be found in Section~\ref{subsec:ILV2}. 
If $\frac{K}{E} \leq \frac{7}{16}$ and hence the mother polar code has to be punctured, additional indices are included in the frozen set such that $\{0,\dots,T\} \subset \bar{Q}_F^N$, with 
\begin{equation}
\label{eq:T}
T = \begin{cases} 
  \left\lceil \frac{3}{4}N - \frac{E}{2} \right\rceil - 1 & \mbox{if } E \geq \frac{3}{4}N  \\ 
  \left\lceil \frac{9}{16}N - \frac{E}{4} \right\rceil - 1 & \text{otherwise} 
  \end{cases}
\end{equation}
This extra freezing is necessary to prevent bits in the information set to become incapable due to puncturing. 
Finally, new indices are added to the frozen set from the universal reliability sequence, starting from the least reliable, until $|\bar{Q}_F^N| = N - K'$. 
To summarize, the frozen set $\bar{Q}_F^N$ is designed in three steps:
\begin{enumerate}
\item Pre-freezing: $Q_1= \{ J(\gamma),\dots, J(\gamma + U - 1) \}$ where $\gamma = 0$ if $\frac{K}{E} \leq \frac{7}{16}$ and $\gamma = E$ otherwise. 
\item Extra freezing: $Q_2 = \{ 0,\dots, T \}$ where $T$ is calculated according to \eqref{eq:T} if $\frac{K}{E} \leq \frac{7}{16}$, otherwise $Q_2 = \emptyset$. 
\item Reliability freezing: $Q_3$ contains the first $N - K' - |Q_1 \cup Q_2|$ elements of $Q_0^{N_{max}-1}$ smaller than $N$ not already included in $Q_1 \cup Q_2$. 
\end{enumerate}
The frozen set is then $\bar{Q}_F^N = Q_1 \cup Q_2 \cup Q_3$; the corresponding bits of $\mathbf{u}$ are set to zero, i.e. $u_i = 0$ for all $i \in \bar{Q}_F^N$. 

\subsubsection{Subchannel allocation}
The information set $\bar{Q}_I^N$ is calculated as the complement of $\bar{Q}_F^N$, and contains $K' = K + n_{PC}$ elements, corresponding to the bit indices that will contain the message bits and the parity check (PC) bits. 
The subchannels to be assigned to PC bits are calculated according to two different strategies: $n_{PC}^{wm}$ bits are selected subject to the weight of the rows of the generator matrix, while $n_{PC}^{lr} = n_{PC} - n_{PC}^{wm}$ are bound to the subchannel reliability.
The set of the PC indices is called $Q_{PC}^N$, with $Q_{PC}^N \subset \bar{Q}_I^N$. 
To begin with, $n_{PC}^{lr}$ bit indices are initially selected as the $n_{PC}^{lr}$ least reliable subchannels in $\bar{Q}_I^N$. 
The index of the remaining $n_{PC}^{wm}$ PC bit is selected as the subchannels corresponding to the row of minimum weight in the transformation matrix among the $K$ most reliable bit indices in $\bar{Q}_I^N$. 
In the case of uncertainty due to the presence of too many rows with the same weight, the index with the highest reliability is selected. 
The row weight $w(g_i)$ of subchannel $i$ corresponds to the number of ones of the $i$-th row $g_i$ of the transformation matrix $\mathbf{G_N}$, and it can be easily calculated as $w(g_i) = 2^{o_i}$, where $o_i$ denotes the number of ones in the binary expansion of $i$ \cite{finite_polar}. 
After the subchannels have been allocated, the $K$ message bits are stored in vector $\mathbf{u}$, i.e. the message is stored in the $K$ indices of $\bar{Q}_I^N \backslash Q_{PC}^N$, and the values of the remaining $n_{PC}$ indices are calculated. 

\subsubsection{PC bit calculation} \label{subsubsec:PC}
The calculation of the PC bits is performed through a cyclic shift register of length 5, initialized to 0. 
Each PC bit is calculated as the XOR of the message bits assigned to preceding subchannels, modulo 5, excluding the previously calculated parity check bits. 
To summarize, a PC bit $u_i$, with $i \in Q_{PC}^N$, is calculated as
\begin{equation}
u_i = \bigoplus_{j = \lfloor i_{PC}/5 \rfloor }^{q-1} u_{5j + p} ~,
\end{equation}
where $q=\lfloor i/5 \rfloor$, $p=i \modulo 5$ and $i_{PC}\in Q_{PC}^N$ is the highest index smaller than $i$ for which $i_{PC} \modulo 5 = p$. If no such index exists, $i_{PC}=0$.

\subsection{Encoding} \label{subsec:enc}
The encoding is performed by the multiplication in $\mathbb{F}_2$ 
\begin{equation}
\mathbf{d} = \mathbf{u} \cdot \mathbf{G_N},
\end{equation} 
where $\mathbf{G_N} = \mathbf{G_2}^{\otimes n}$, with $\mathbf{G_2} =  \left[\begin{smallmatrix} 1&0\\1&1 \end{smallmatrix}\right]$. 
Encoding complexity can be proved to be $O(N \log_2(N))$ \cite{arikan}. 
However, the recursive structure of the transformation matrix suggests the possibility to have parallel implementation \cite{syst_polar}. 
If $N/2$ processing units are available, the encoding latency can be reduced to $O(\log_2(N))$; a tradeoff between hardware complexity and encoding latency can be found in between these extremes. 

\subsection{Sub-block interleaver} \label{subsec:ILV2}
The $N$ encoded bits are then interleaved before performing the rate matching. 
This interleaver $J$ divides the $N$ encoded bits stored in $\mathbf{d}$ into 32 blocks of length $B = \frac{N}{32}$ bits, interleaving the blocks according to a list of 32 integers $P$ and obtaining the vector $\mathbf{y}$ as illustrated in Figure~\ref{fig:SB_inter}. 
Encoded bit $d_i$ is assigned to interleaved bit $y_j$, where
\begin{equation}
i = J(j) = B \cdot P(\lfloor j/B \rfloor) + q,
\end{equation}
and $q = j \modulo B$, for all $j = 0,\dots,N-1$. 

\subsection{Rate matching} \label{subsec:RM}
\begin{figure}
  \centering
  \includegraphics[width=0.45\textwidth]{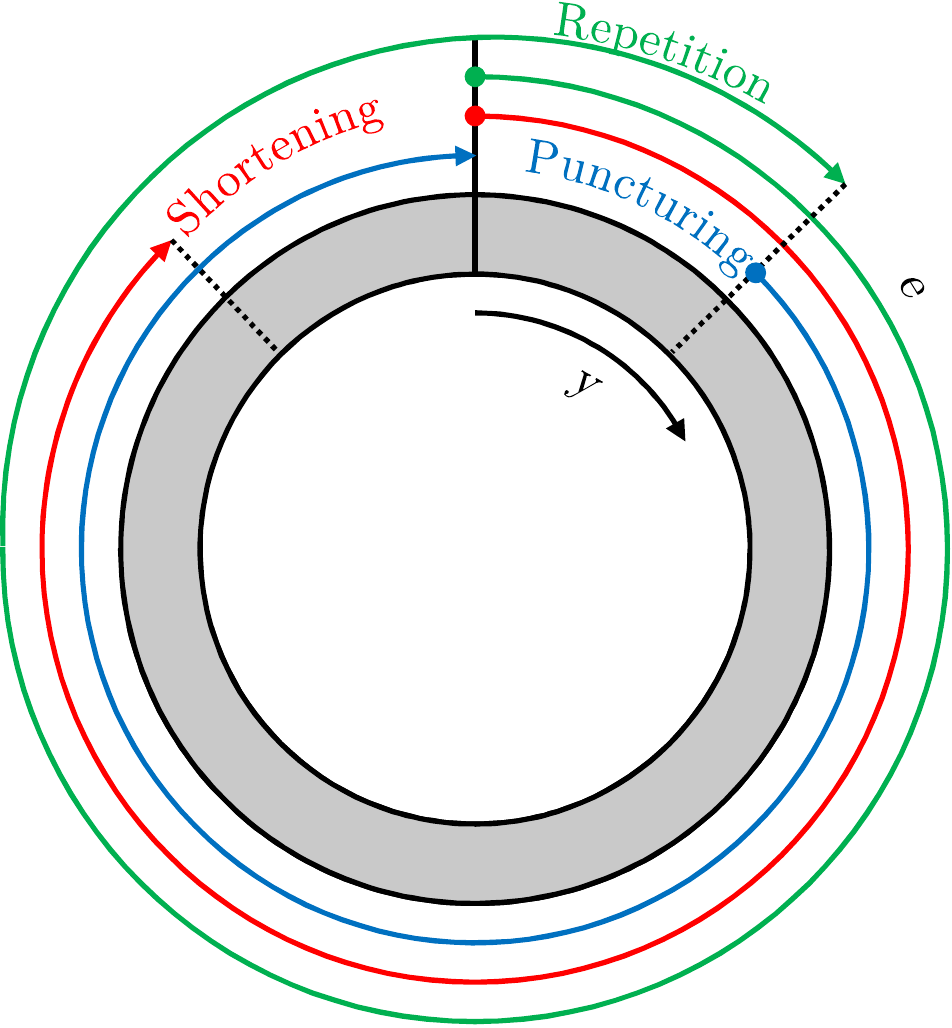}
  \caption{Circular buffer design for rate-matching.}
  \label{fig:RM_buffer}
\end{figure}
Rate matching is performed by a circular buffer, and the codeword $\mathbf{e}$ of length $E$ bits is calculated.
As previously mentioned, three possible rate-matching schemes are foreseen:
\begin{itemize}
\item Puncturing: if $E \leq N$ and $R\le\frac{7}{16}$, the mother code is punctured. In this case, the first $U = N - E$ bits are not transmitted, hence $e_i = y_{i + U}$ for $i = 0,\dots,E-1$. 
\item Shortening: if $E \leq N$ and $R>\frac{7}{16}$, the mother code is shortened. In this case, the last $U = N - E$ bits are not transmitted, hence $e_i = y_i$ for $i = 0,\dots,E-1$. 
\item Repetition: if $E > N$, the first $U = N - E$ bits are transmitted twice, with $e_i = y_{i \modulo N}$ for $i = 0,\dots,E-1$. 
\end{itemize}
The operating principle of the rate-matcher based on the circular buffer is illustrated in Figure~\ref{fig:RM_buffer}. 

\subsection{Channel interleaver} \label{subsec:ILV3}

Before passing the rate-matched codeword to the modulator, the bits in $\mathbf{e}$ are interleaved one more time using a triangular bit interleaver. 
This interleaver has been considered necessary to improve the coding performance of the coding scheme for high-order modulation; it is not applied for every use case, hence it is triggered by a parameter $I_{BIL}$.  
In particular, the channel interleaver is activated for PUCCH and PUSCH UCIs ($I_{BIL}=1$), while it is bypassed in the case of PBCH payloads and PDCCH DCIs ($I_{BIL}=0$).

\begin{figure}
  \centering
  \includegraphics[width=0.45\textwidth]{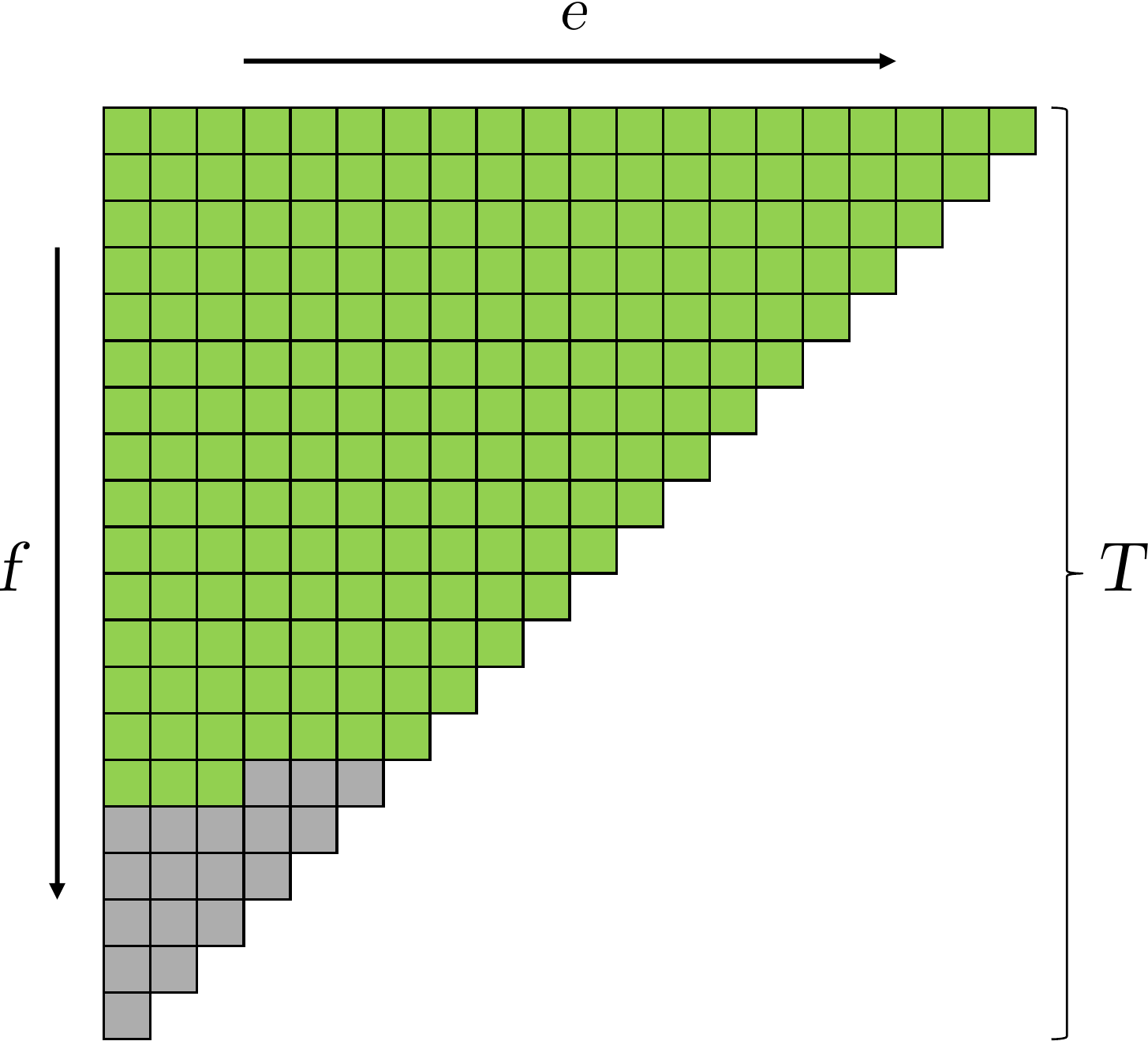}
  \caption{Channel interleaver design.}
  \label{fig:CH_inter}
\end{figure}

The channel interleaver is formed by an isosceles triangular structure of length $T$ bits, where $T$ is the smallest integer such that $\frac{T (T + 1)}{2} \geq E$; its value can be be calculated as $T = \left\lceil \frac{\sqrt{8E+1}-1}{2} \right\rceil$. 
The encoded bits in $\mathbf{e}$ are written into the rows of the triangular structure, while the interleaved vector $\mathbf{f}$ is obtained by reading bits out of the structure in columns. 
The construction of the interleaving pattern is illustrated in Figure~\ref{fig:CH_inter}. 
In more detail, an auxiliary $T \times T$ matrix $\mathbf{V}$ is created on the basis of $\mathbf{e}$ with 
\begin{equation}
\label{eq:V}
V_{i,j} = \begin{cases} 
  \text{NULL} & \mbox{if } i+j \geq T \mbox{ or }  r(i)+j \geq E \\
  e_{r(i)+j} & \text{otherwise} 
  \end{cases}
\end{equation}
where $r(i) = \frac{i(2T-i+1)}{2}$. 
The interleaved vector $\mathbf{f}$ is created by appending the columns of $\mathbf{V}$ while skipping the NULL entries. 
This triangular interleaver has been proposed in 5G standardization because of the practical advantages provided by its high parallelism factor, due to its maximum contention-free property, and its high flexibility. 

\subsection{Block concatenation} 
\label{subsec:conc}
If segmentation has been activated at the beginning of the process, the two codewords of length $E$ are appended in order to obtain a unique block of length $G$. 
If $G = 2E+1$, a zero bit is appended at the end of the second codeword. 

\section{Decoding considerations} \label{sec:dec}
\begin{figure*}
\begin{subfigure}{.5\textwidth}
  \centering
  \includegraphics[width=0.95\textwidth]{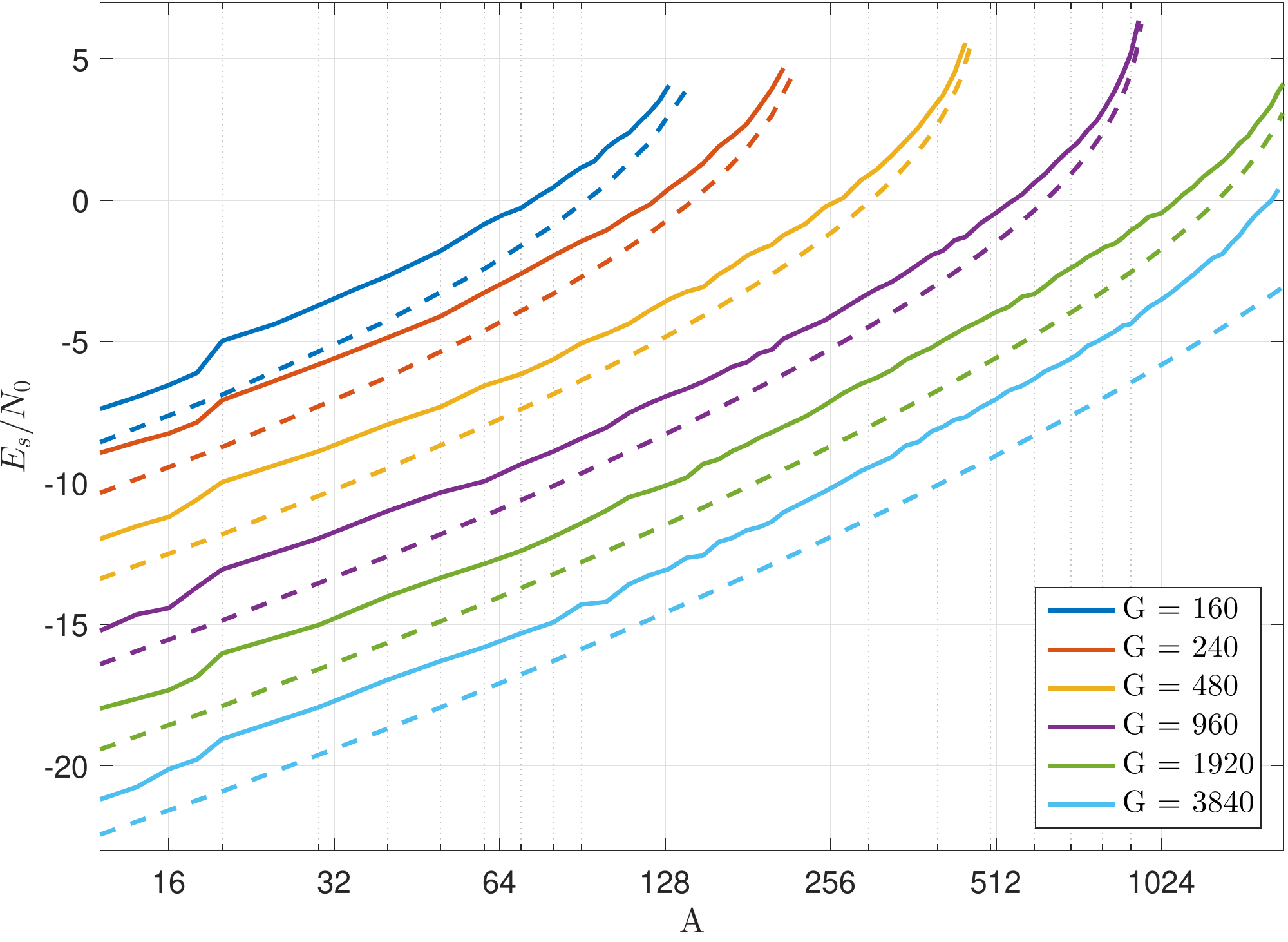}
  \caption{PUCCH}
  \label{fig:perf_PUCCH}
\end{subfigure}
\begin{subfigure}{.5\textwidth}
  \centering
  \includegraphics[width=0.95\textwidth]{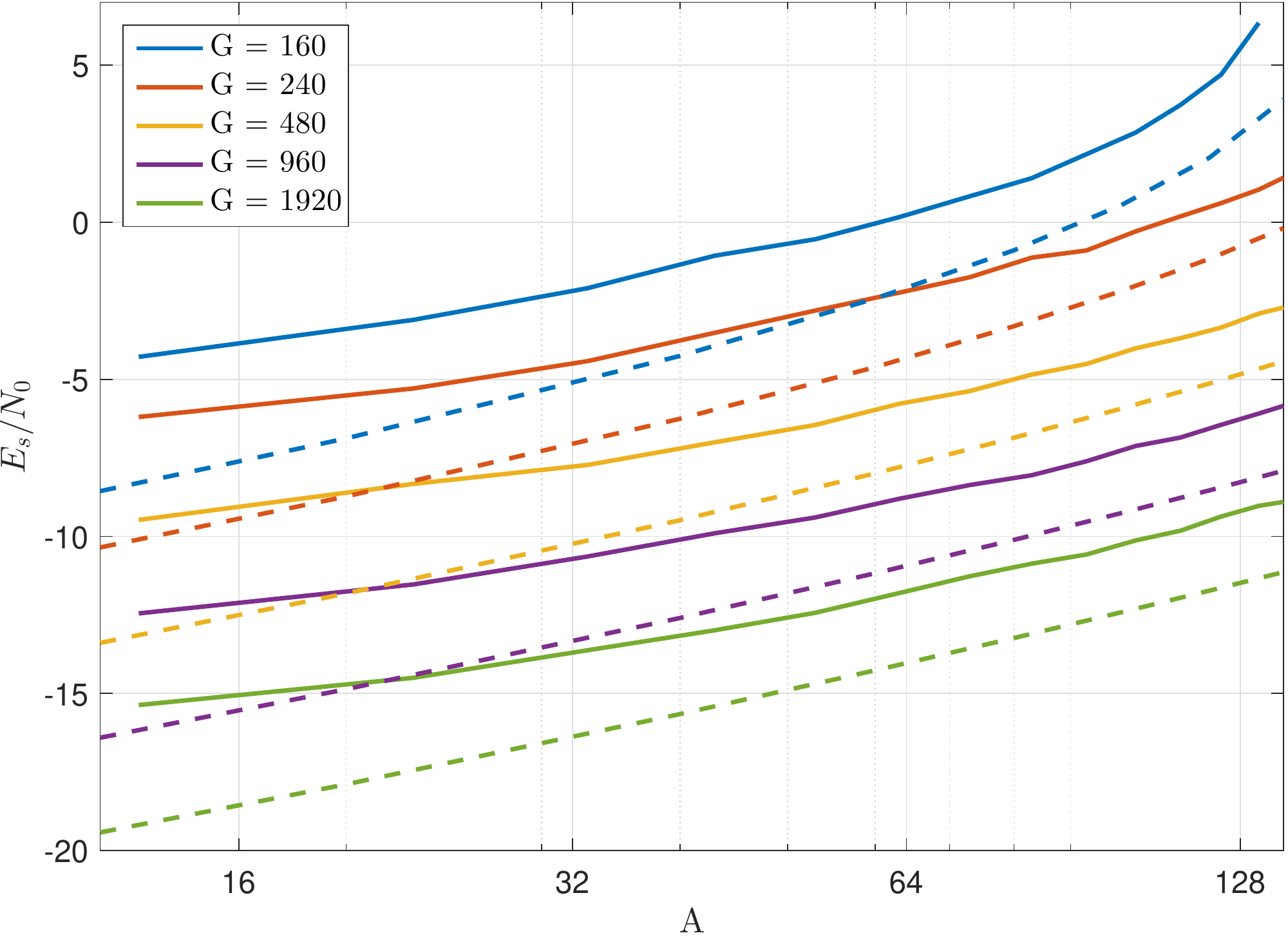}
  \caption{PDCCH}
  \label{fig:perf_PDCCH}
\end{subfigure}
\caption{SNR required to achieve BLER of $10^{-3}$.}
\label{fig:perf}
\end{figure*}

Even if 3GPP does not provide any suggestion on the decoding, the final code structure gives to the reader some guidance on decoding strategies for polar codes standardized in 5G. 
Decoding can be easily performed inverting the encoding chain presented in Figure~\ref{fig:5Gchain}. 
The received encoded symbols, after a further block segmentation and deinterleaving step for the uplink, are padded and deinterleaved to reach the mother polar code length. 
The nature of the padding depends on the rate matching strategy employed; zeros are appended in the case of puncturing, saturated symbols for shortening, while repeated symbols are combined in the case of repetition. 
After padding, the number of coded symbols is a power of two, so they can be decoded through common polar code decoders \cite{arikan,tal_list,BP_pc,SCAN_pc}; the handling of the assistant bits included in the code is crucial in this phase and will be discussed in detail. 
In the following, we delve into SCL-based decoding of polar codes, which is today the de-facto standard decoder for these codes, describing how this algorithm takes advantage of assistant bits in several ways. 
Then, bits are possibly deinterleaved and concatenated according to code and channel parameters before being passed to the upper communication layer.  

Assistant bits in CA-SCL decoders can be used in three different ways: $(i)$ they can provide an early termination mechanism, or $(ii)$ reduce the FAR by decreasing undetected errors probability, or $(iii)$ improve the overall BLER of the code. 
Two different kind of assistant bits, namely CRC and PC bits, were originally introduced to be used differently during the decoding: distributed CRC bits targeted early termination and FAR mitigation, while PC bits were designed to improve the error correction performance of the code. 
Eventually, this distinction vanished during standard development, leaving the management of these bits to the designer of the decoder. 

The input bits interleaver introduced for DCI induces a distributed CRC to the polar codeword, whose bits become analogous to the PC bits present in the UCI. 
The presence of these checks along the decoding can be used either for early termination, namely stopping the decoding if the check is not passed by any of the paths of SCL, or to improve BLER if used as dynamically frozen bits \cite{dynFroz}. 
CRC bits at the end of the codeword can be either used the reduce the FAR or to improve error correction capability of the code. 
A decoder can trade off among these effects depending on current requirements. 

More in detail, the early decoding termination property is triggered by the failure of the check for all active paths within the list decoder; the decoding process continues when at least one path passes the check. 
However, how a decoder should manage failing paths may alter BLER performance of the code \cite{WCL}. 
The straightforward approach is to maintain the failing paths in the list to keep the number of active paths constant, simplifying the decoder implementation. 
An alternative is to deactivate failed paths immediately after each bit estimation. 
This improves BLER performance while reducing the computational complexity and power consumption of the decoder, however making the number of surviving paths varying \cite{PC_hua}. 
Finally, distributed assistant bits can be considered as dynamically frozen bits, assigning to the bit the value imposed by the check. 
This technique and the failed path deactivation yield comparable BLER performance. 
As dynamically frozen bits, however, assistant bits do not improve the computational complexity and power consumption of the decoder, and prevent early-termination since all surviving paths are guaranteed to pass the check \cite{WCL}. 

Figure~\ref{fig:perf} shows the Signal-to-Noise Ratio (SNR) $E_s/N_0$ required by the 5G polar code to achieve a BLER of $10^{-3}$ as a function of the message length $A$ for different payload lengths $G$. 
Binary phase-shift keying (BPSK) modulation over AWGN channel has been used for the simulations. 
Polar codes are decoded through SCL decoding with a list size of $8$, as proposed by 3GPP for baseline. 
Dotted lines represent $O(n^{-2})$ approximation of the Polyanskyi-Poor-Verd\`u (PPV) meta-converse bound for the finite blocklength regime \cite{PPV_approx}. 
Figure~\ref{fig:perf_PUCCH} considers the PUCCH case, where PC bits introduced for $A \le 19$ are considered as dynamic frozen bits. 
Their presence brings a noticeable performance improvement, observed by the sudden jump in the BLER curves at $A>19$.
Figure~\ref{fig:perf_PDCCH} presents the PDCCH case, where early termination is enabled and the SCL decoder maintains failing paths in the list.
The BLER performance is degraded compared to PUCCH due to the larger CRC inserted, however with a better FAR mitigation and a more efficient early termination mechanism. 

\section{Conclusion} \label{sec:conc}
In this work, we have detailed the polar code encoding process within the $5^{\text{th}}$ generation wireless systems standard, providing the reader with a user-friendly description to understand, implement and simulate 5G-compliant polar code encoding.  
This encoding chain showcases the successful efforts of the 3GPP standardization body to meet the various requirements on the code for the eMBB control channel:  low description complexity and low encoding complexity, while covering a wide range of code lengths and code rates.  
Throughout this work, we hinted that the standardization process also took the receiver side into account.  
Typical for modern channel coding, the encoder was designed such that the decoder can be implemented with feasible complexity and operate at the required latency, assuming state-of-the-art decoders and hardware. 
New decoding principles or decoding architectures, however, may now be developed to optimize decoding complexity or improve error-rate performance.

With the 5G eMBB control channel, polar codes have found their first adoption into a standard only 10 years after their invention.  
This standardization has triggered further academic and industrial research into polar coding, and adoption in future standards and systems can be foreseen, given the flexibility of code and decoder design that polar codes offer.  
Our detailed description of the 5G polar codes, including the individual components of the encoding chain, may serve as a reference to further development of polar codes. 


\bibliographystyle{IEEEbib}
\bibliography{IEEEabrv,refs}

\end{document}

%% file: figures/G_2.tex
\begin{tikzpicture}

\draw  (-2.5,0.5) edge (2.5,0.5) ;
\draw  (-2.5,-1) edge (2.5,-1);
\draw  (0,0.5) ellipse (.5 and .5);
\draw  (0,1) edge (0,-1);
\node at (-3,0.5) {\huge $u_0$};
\node at (-3,-1) {\huge $u_1$};
\node at (4.75,0.5) {\huge $d_0=u_0 \oplus u_1$};
\node at (4,-1) {\huge $d_1=u_1$};
\draw [fill=white] (0,-1) ellipse (.1 and .1);
\end{tikzpicture}

%% file: figures/G_8.tex
\begin{tikzpicture}

\draw  (-1.5,0.5) -- (12.5,0.5) ;
\draw  (-1.5,-1) -- (12.5,-1);
\draw  (0.5,0.5) ellipse (.5 and .5);
\draw  (0.5,1) -- (0.5,-1);
\node at (-6,0.5) {\huge $\delta_0=0.99$};
\node at (-6,-1) {\huge $\delta_1=0.88$};
\node at (-6,-2.5) {\huge $\delta_2=0.81$};
\node at (-6,-4) {\huge $\delta_3=0.32$};
\node at (-6,-5.5) {\huge $\delta_4=0.68$};
\node at (-6,-7) {\huge $\delta_5=0.19$};
\node at (-6,-8.5) {\huge $\delta_6=0.12$};
\node at (-6,-10) {\huge $\delta_7=0.01$};
\node [text=blue] at (-3,0.5) {\huge $u_0=0$};
\node [text=blue] at (-3,-1) {\huge $u_1=0$};
\node [text=blue] at (-3,-2.5) {\huge $u_2=0$};
\node at (-3,-4) {\huge $u_3=1$};
\node [text=blue] at (-3,-5.5) {\huge $u_4=0$};
\node at (-3,-7) {\huge $u_5=0$};
\node at (-3,-8.5) {\huge $u_6=1$};
\node at (-3,-10) {\huge $u_7=1$};
\node at (-1,1) {\large $0$};
\node at (-1,-0.5) {\large $0$};
\node at (-1,-2) {\large $0$};
\node at (-1,-3.5) {\large $1$};
\node at (-1,-5) {\large $0$};
\node at (-1,-6.5) {\large $0$};
\node at (-1,-8) {\large $1$};
\node at (-1,-9.5) {\large $1$};
\node at (2,1) {\large $0$};
\node at (2,-0.5) {\large $0$};
\node at (2,-2) {\large $1$};
\node at (2,-3.5) {\large $1$};
\node at (2,-5) {\large $0$};
\node at (2,-6.5) {\large $0$};
\node at (2,-8) {\large $0$};
\node at (2,-9.5) {\large $1$};
\node at (6,1) {\large $1$};
\node at (6,-0.5) {\large $1$};
\node at (6,-2) {\large $1$};
\node at (6,-3.5) {\large $1$};
\node at (6,-5) {\large $0$};
\node at (6,-6.5) {\large $1$};
\node at (6,-8) {\large $0$};
\node at (6,-9.5) {\large $1$};
\node at (12,1) {\large $1$};
\node at (12,-0.5) {\large $0$};
\node at (12,-2) {\large $1$};
\node at (12,-3.5) {\large $0$};
\node at (12,-5) {\large $0$};
\node at (12,-6.5) {\large $1$};
\node at (12,-8) {\large $0$};
\node at (12,-9.5) {\large $1$};
\node at (14,0.5) {\huge $d_0=1$};
\node at (14,-1) {\huge $d_1=0$};
\node at (14,-2.5) {\huge $d_2=1$};
\node at (14,-4) {\huge $d_3=0$};
\node at (14,-5.5) {\huge $d_4=0$};
\node at (14,-7) {\huge $d_5=1$};
\node at (14,-8.5) {\huge $d_6=0$};
\node at (14,-10) {\huge $d_7=1$};

\node at (0.5,-11) {\huge step 1};
\node at (4,-11) {\huge step 2};
\node at (9,-11) {\huge step 3};

\draw (-1.5,-4) -- (12.5,-4);
\draw (-1.5,-2.5) -- (12.5,-2.5);
\draw (-1.5,-5.5) -- (12.5,-5.5);
\draw (-1.5,-7) -- (12.5,-7);
\draw (-1.5,-8.5) -- (12.5,-8.5);
\draw (-1.5,-10) -- (12.5,-10);
\draw (0.5,-4) -- (0.5,-2);
\draw (0.5,-7) -- (0.5,-5);
\draw (0.5,-10) -- (0.5,-8);
\draw  (0.5,-2.5) ellipse (0.5 and 0.5);
\draw  (0.5,-5.5) ellipse (0.5 and 0.5);
\draw  (0.5,-8.5) ellipse (0.5 and 0.5);
\draw (3.5,-4) -- (3.5,-0.5);
\draw (4.5,-2.5) -- (4.5,1) node (v1) {};
\draw  (3.5,-1) ellipse (0.5 and 0.5);
\draw  (4.5,0.5) ellipse (0.5 and 0.5);
\draw (3.5,-6.5) -- (3.5,-10);
\draw (4.5,-5) -- (4.5,-8.5);
\draw  (4.5,-5.5) ellipse (0.5 and 0.5);
\draw  (3.5,-7) ellipse (0.5 and 0.5);
\draw (7.5,-10) -- (7.5,-3.5);
\draw (8.5,-8.5) -- (8.5,-2);
\draw (9.5,-7) -- (9.5,-0.5);
\draw (10.5,-5.5) -- (10.5,1);
\draw [fill=white] (0.5,-10) ellipse (.1 and .1);
\draw [fill=white] (0.5,-1) ellipse (.1 and .1);
\draw [fill=white] (0.5,-4) ellipse (.1 and .1);
\draw [fill=white] (0.5,-7) ellipse (.1 and .1);
\draw [fill=white] (3.5,-10) ellipse (.1 and .1);
\draw [fill=white] (4.5,-2.5) ellipse (.1 and .1);
\draw [fill=white] (3.5,-4) ellipse (.1 and .1);
\draw [fill=white] (4.5,-8.5) ellipse (.1 and .1);
\draw [fill=white] (0.5,-7) ellipse (.1 and .1);
\draw [fill=white] (7.5,-10) ellipse (.1 and .1);
\draw [fill=white] (10.5,-5.5) ellipse (.1 and .1);
\draw [fill=white] (9.5,-7) ellipse (.1 and .1);
\draw [fill=white] (8.5,-8.5) ellipse (.1 and .1);
\draw  (7.5,-4) ellipse (0.5 and 0.5);
\draw  (8.5,-2.5) ellipse (0.5 and 0.5);
\draw  (9.5,-1) ellipse (0.5 and 0.5);
\draw  (10.5,0.5) ellipse (0.5 and 0.5);

\draw  (-0.5,1.5) rectangle (1.5,-10.5);
\draw  (2.5,1.5) rectangle (5.5,-10.5);
\draw  (6.5,-10.5) rectangle (11.5,1.5);
\end{tikzpicture}

%% file: figures/sc-dec.tex
\begin{tikzpicture}[scale=1.9, thick]
\newcommand\Triangle[1]{-- ++(0:2*#1) -- ++(120:2*#1) --cycle}
\newcommand\Square[1]{+(-#1,-#1) rectangle +(#1,#1)}

  \fill [gray, very thick] (0,0) circle [radius=.05];
  
  \fill [gray, very thick] (-2,-0.5) circle [radius=.05]; 
  \fill [gray, very thick] (2,-0.5) circle [radius=.05]; 

  \fill [gray, very thick] (-3,-1) circle [radius=.05];
  \fill [gray, very thick] (-1,-1) circle [radius=.05];
  \fill [gray, very thick] (1,-1) circle [radius=.05];
  \fill [gray, very thick] (3,-1) circle [radius=.05];

  \draw (-3.5,-1.5) circle [radius=.05];
  \draw (-2.5,-1.5) circle [radius=.05];
  \draw (-1.5,-1.5) circle [radius=.05];
  \fill (-0.5,-1.5) circle [radius=.05];
  \draw (0.5,-1.5) circle [radius=.05];
  \fill (1.5,-1.5) circle [radius=.05];
  \fill (2.5,-1.5) circle [radius=.05];
  \fill (3.5,-1.5) circle [radius=.05];

  \node at (-3.5,-1.7) {$\hat{u}_0=0$};
  \node at (-2.5,-1.7) {$\hat{u}_1=0$};
  \node at (-1.5,-1.7) {$\hat{u}_2=0$};
  \node at (-0.5,-1.7) {$\hat{u}_3=1$};
  \node at (0.5,-1.7) {$\hat{u}_4=0$};
  \node at (1.5,-1.7) {$\hat{u}_5=0$};
  \node at (2.5,-1.7) {$\hat{u}_6=1$};
  \node at (3.5,-1.7) {$\hat{u}_7=1$};

  \draw (0,-0.05) -- (-2,-0.45);
  \draw (0,-0.05) -- (2,-0.45);

  \draw (-2,-0.55) -- (-3,-0.95);
  \draw (-2,-0.55) -- (-1,-0.95);
  \draw (2,-0.55) -- (1,-0.95);
  \draw (2,-0.55) -- (3,-0.95);

  \draw (-3,-1.05) -- (-3.5,-1.45);
  \draw (-3,-1.05) -- (-2.5,-1.45);
  \draw (-1,-1.05) -- (-1.5,-1.45);
  \draw (-1,-1.05) -- (-0.5,-1.45);
  \draw (1,-1.05) -- (0.5,-1.45);
  \draw (1,-1.05) -- (1.5,-1.45);
  \draw (3,-1.05) -- (2.5,-1.45);
  \draw (3,-1.05) -- (3.5,-1.45);

  \draw [very thin,gray,dashed] (-4,0) -- (4,0);
  \draw [very thin,gray,dashed] (-4,-0.5) -- (4,-0.5);
  \draw [very thin,gray,dashed] (-4,-1) -- (4,-1);
  \draw [very thin,gray,dashed] (-4,-1.5) -- (4,-1.5);

  \node at (-4.25,0) {$t=3$};
  \node at (-4.25,-0.5) {$t=2$};
  \node at (-4.25,-1) {$t=1$};
  \node at (-4.25,-1.5) {$t=0$};
  
  \draw [->] (0,0.5) -- (0,0.07) node [right=-.1cm,midway] {$[e,0,e,0,e,1,0,1]$};  

  \draw [->] (-.12,-.03) -- (-1.88,-0.38) node [above=-.08cm,midway,rotate=11] {$[e,1,e,1]$};
  \draw [->] (-1.88,-0.48) -- (-.12,-0.12) node [below=-.08cm,midway,rotate=11] {$[1,1,1,1]$};
  \draw [<-] (1.88,-0.48) -- (.12,-0.12) node [below=-.08cm,midway,rotate=-11] {$[e,1,0,1]$};

  \draw [->] (-2.06,-0.52) -- (-2.94,-0.87) node [above=-.1cm,midway,rotate=21] {$[e,0]$};
  \draw [->] (-2.94,-0.98) -- (-2.06,-0.63) node [below=-.1cm,midway,rotate=21] {$[0,0]$};
  \draw [<-] (-1.94,-0.52) -- (-1.06,-0.87) node [above=-.1cm,midway,rotate=-21] {$[1,1]$};
  \draw [<-] (-1.06,-0.98) -- (-1.94,-0.63) node [below=-.1cm,midway,rotate=-21] {$[e,1]$};
  \draw [<-] (2.94,-0.98) -- (2.06,-0.63) node [below=-.1cm,midway,rotate=-21] {$[0,1]$};
  \draw [->] (1.94,-0.52) -- (1.06,-0.87) node [above=-.1cm,midway,rotate=21] {$[e,0]$};
  \draw [->] (1.06,-0.98) -- (1.94,-0.63) node [below=-.1cm,midway,rotate=21] {$[0,0]$};

  \draw [->] (-3.06,-1.04) -- (-3.43,-1.34) node [above=-.1cm,midway,rotate=37] {$[e]$};
  \draw [->] (-3.43,-1.46) -- (-3.06,-1.16) node [below=-.1cm,midway,rotate=37] {$[0]$};
  \draw [<-] (-2.94,-1.04) -- (-2.57,-1.34) node [above=-.1cm,midway,rotate=-37] {$[0]$};
  \draw [<-] (-2.57,-1.46) -- (-2.94,-1.16) node [below=-.1cm,midway,rotate=-37] {$[0]$};
  \draw [->] (-1.06,-1.04) -- (-1.43,-1.34) node [above=-.1cm,midway,rotate=37] {$[e]$};
  \draw [->] (-1.43,-1.46) -- (-1.06,-1.16) node [below=-.1cm,midway,rotate=37] {$[0]$};
  \draw [<-] (-0.94,-1.04) -- (-0.57,-1.34) node [above=-.1cm,midway,rotate=-37] {$[1]$};
  \draw [<-] (-0.57,-1.46) -- (-0.94,-1.16) node [below=-.1cm,midway,rotate=-37] {$[1]$};  
  \draw [<-] (3.43,-1.46) -- (3.06,-1.16) node [below=-.1cm,midway,rotate=-37] {$[1]$};
  \draw [->] (2.94,-1.04) -- (2.57,-1.34) node [above=-.1cm,midway,rotate=37] {$[1]$};
  \draw [->] (2.57,-1.46) -- (2.94,-1.16) node [below=-.1cm,midway,rotate=37] {$[1]$};
  \draw [<-] (1.06,-1.04) -- (1.43,-1.34) node [above=-.1cm,midway,rotate=-37] {$[0]$};
  \draw [<-] (1.43,-1.46) -- (1.06,-1.16) node [below=-.1cm,midway,rotate=-37] {$[0]$};
  \draw [->] (0.94,-1.04) -- (0.57,-1.34) node [above=-.1cm,midway,rotate=37] {$[e]$};
  \draw [->] (0.57,-1.46) -- (0.94,-1.16) node [below=-.1cm,midway,rotate=37] {$[0]$};  

\end{tikzpicture}

%% file: figures/sc-node.tex
\begin{tikzpicture}[scale=1.9, thick]
\newcommand\Triangle[1]{-- ++(0:2*#1) -- ++(120:2*#1) --cycle}
\newcommand\Square[1]{+(-#1,-#1) rectangle +(#1,#1)}

  \fill [gray, very thick] (-1,0) circle [radius=.05];
  
  
  \draw [very thin,gray,dashed] (-2,0) -- (0.1,0);
  \draw [very thin,gray,dashed] (-2,-.5) -- (0.1,-0.5);
  \draw [very thin,gray,dashed] (-2,-1) -- (0.1,-1);

  \fill [red, very thick] (-1.0,-.5) circle [radius=.05]; 

  \fill [gray, very thick] (-1.5,-1) circle [radius=.05];
  \fill [gray, very thick] (-.5,-1.0) circle [radius=.05];

  \draw (-1,-0.05) -- (-1,-.45);
  
  \draw (-1,-.55) -- (-1.5,-.95);
  \draw (-1,-.55) -- (-.5,-.95);

  \node at (-2.25,0) {$t+1$};
  \node at (-2.25,-.5) {$t$};
  \node at (-2.25,-1) {$t-1$};

  \draw [->] (-1.05,-0.05) -- (-1.05,-.45) node [left=-.08cm,midway] {$\bm{\alpha}$};
  \draw [->] (-.95,-.45) -- (-0.95,-0.05) node [right=-.08cm,midway] {$\bm{\beta}$};

  \draw [->] (-1.06,-.55) -- (-1.5,-.9) node [above=-.1cm,midway,rotate=40] {$\bm{\alpha}^{\ell}$};
  \draw [->] (-1.44,-.95) -- (-1.0,-0.6) node [below=-.1cm,midway,rotate=40] {$\bm{\beta}^{\ell}$};

  \draw [<-] (-.94,-.55) -- (-.5,-.9) node [above=-.1cm,midway,rotate=-40] {$\bm{\beta}^{\text{r}}$};
  \draw [<-] (-.56,-.95) -- (-0.975,-.625) node [below=-.1cm,midway,rotate=-40] {$\bm{\alpha}^{\text{r}}$};

\end{tikzpicture}

%% file: figures/5Gchain.tikz
\begin{tikzpicture}[scale=1, thick]

\draw [->, very thick] (-5.5,-2.25) -- ++(1.5,0) {};
\node [] at (-5.25,-1.95) {$\mathbf{a}$}; 
\draw[-] (-4.75,-2.5) -- (-4.5, -2.0);
\node [] at (-4.75,-2.75) {$\mathbf{A}$}; 

\draw [fill=white!50!red] (-4,-1.5) rectangle ++(3,-1.5) node [pos=0.5,align=center] {Code-Block \\ Segmentation};
\draw [->] (-0.25,-2.25) -- ++(-0.75,0) {};
\node [] at (0.25,-2.25) {$I_{seg}$};
\draw [->, very thick] (-2.5,-3) -- ++(0,-1.) {};
\node [] at (-2.25,-3.25){$\mathbf{a'}$};
\draw[-] (-2.75,-3.5) -- (-2.25, -3.75);
\node [] at (-3,-3.5) {$\mathbf{A'}$}; 

\draw [fill=white!50!orange] (-4,-4) rectangle ++(3,-1.5) node [pos=0.5,align=center] {CRC \\ Encoder};
\draw [->] (-0.25,-4.75) -- ++(-0.75,0) {};
\node [] at (0.05,-4.75) {$L$};
\draw [->, very thick] (-2.5,-5.5) -- ++(0,-1) {};
\node [] at (-2.25,-5.75) {$\mathbf{c}$};
\draw[-] (-2.75,-6) -- (-2.25, -6.25);
\node [] at (-3,-6) {$\mathbf{K}$}; 

\draw [fill=white!50!yellow] (-4,-6.5) rectangle ++(3,-1.5) node [pos=0.5,align=center] {Input Bits \\ Interleaver};
\draw [->] (-4.75,-7.25) -- ++(0.75,0) {};
\node [] at (-5.15,-7.25) {$I_{IL}$};
\draw [->, very thick] (-1,-7.25) -- ++(2.5,0) {};
\node [] at (-0.5,-7) {$\mathbf{c'}$};
\draw[-] (0.25,-7.5) -- (0.5,-7);
\node [] at (0.25,-7.75) {$\mathbf{K}$}; 

\draw [fill=white!50!orange] (1.5,-6.5) rectangle ++(3,-1.5) node [pos=0.5,align=center] {Subchannel \\ Allocation};
\draw [->, very thick] (3,-6.5) -- ++(0,1) {};
\node [] at (3.25,-6.25) {$\mathbf{u}$};
\draw[-] (2.75,-6) -- (3.25, -5.75);
\node [] at (2.5,-6) {$\mathbf{N}$}; 

\draw [fill=white!50!orange] (1.5,-4) rectangle ++(3,-1.5) node [pos=0.5,align=center] {Polar Code \\ Encoding};
\draw [->, very thick] (3,-4) -- ++(0,1) {};
\node [] at (3.25,-3.75) {$\mathbf{d}$};
\draw[-] (2.75,-3.5) -- (3.25, -3.25);
\node [] at (2.5,-3.5) {$\mathbf{N}$}; 

\draw [fill=white!50!orange] (1.5,-1.5) rectangle ++(3,-1.5) node [pos=0.5,align=center] {Sub-Block \\ Interleaver};
\draw [->, very thick] (4.5,-2.25) -- ++(2.5,0) {};
\node [] at (5,-2) {$\mathbf{y}$};
\draw[-] (5.75,-2.5) -- (6, -2);
\node [] at (5.75,-2.75) {$\mathbf{N}$}; 

\draw [fill=white!50!orange] (7,-1.5) rectangle ++(3,-1.5) node [pos=0.5,align=center] {Rate Matching \\ Circular Buffer };
\draw [->, very thick] (8.5,-3) -- ++(0,-1) {};
\node [] at (8.75,-3.25) {$\mathbf{e}$};
\draw[-] (8.25,-3.5) -- (8.75, -3.75);
\node [] at (8,-3.5) {$\mathbf{E}$}; 

\draw [fill=white!50!red] (7,-4) rectangle ++(3,-1.5) node [pos=0.5,align=center] {Channel \\ Interleaver };
\draw [->] (6.25,-4.75) -- ++(0.75,0) {};
\node [] at (5.75,-4.75) {$I_{BIL}$};
\draw [->, very thick] (8.5,-5.5) -- ++(0,-1) {};
\node [] at (8.75,-5.75) {$\mathbf{f}$};
\draw[-] (8.25,-6) -- (8.75, -6.25);
\node [] at (8,-6) {$\mathbf{E}$}; 

\draw [fill=white!50!red] (7,-6.5) rectangle ++(3,-1.5) node [pos=0.5,align=center] {Code-Block \\ Concatenation };
\draw [->] (6.25,-7.25) -- ++(0.75,0) {};
\node [] at (5.75,-7.25) {$I_{seg}$};
\draw [->, very thick] (10,-7.25) -- ++(1.5,0) {};
\node [] at (10.25,-7) {$\mathbf{g}$};
\draw[-] (10.75,-7.5) -- (11,-7);
\node [] at (10.75,-7.75) {$\mathbf{G}$};

\end{tikzpicture}